\documentclass[usenatbib, twocolumn, nofootinbib]{aastex631}

\usepackage{multirow}
\usepackage{wrapfig}
\usepackage{fp}
\usepackage{natbib}
\usepackage{color}
\usepackage{amsmath,amsfonts,bm}
\usepackage{hyperref}
\usepackage[T1]{fontenc}
\usepackage{graphicx}	
\usepackage{cleveref}
\usepackage[mathlines]{lineno}
\usepackage{fontawesome}

\newcommand{\sfour}{S4-Wide}
\newcommand{\sfourdeep}{S4-Ultra deep}
\newcommand{\cmbhd}{\mbox{CMB-HD}}
\newcommand{\spt}{SPT}

\newcommand{\act}{ACT}
\newcommand{\planck}{{\it Planck}}

\newcommand{\snr}{S/N}

\newcommand{\fsky}{f_{\rm sky}}
\newcommand{\binnedcluscountsquantities}{z, M_{L}, q}
\newcommand{\binnedcluscountsquantitieswithsubscripts}{z_{i}, M_{L_{j}}, q_{k}}
\newcommand{\binnedcluscounts}{N(\binnedcluscountsquantities)}
\newcommand{\binnedcluscountswithsubscripts}{N(\binnedcluscountsquantitieswithsubscripts)}
\newcommand{\binnedcluscountsnolensing}{N(z, q)}
\newcommand{\summnu}{\sum m_{\nu}}
\newcommand{\wde}{w_{0}}
\newcommand{\taure}{\tau_{\rm re}}
\newcommand{\omchsq}{\Omega_{c}h^{2}}
\newcommand{\ombhsq}{\Omega_{b}h^{2}}
\newcommand{\vireffeta}{\eta_{\rm v}}

\newcommand{\rvir}{R_{\rm 500c}}
\newcommand{\thetavir}{\theta_{500c}}
\newcommand{\mvir}{M_{\rm 500c}}
\newcommand{\Ysz}{Y_{\rm SZ}}
\newcommand{\Yvir}{Y_{{\rm SZ}_{500c}}}

\newcommand{\mev}{{\rm meV}}
\newcommand{\boxsize}{$2^{\circ} \times 2^{\circ}$}
\newcommand{\pixres}{0.^{\prime}5}
\newcommand{\ukam}{\ensuremath{\mu}{\rm K{\text -}arcmin}}
\newcommand{\uk}{\ensuremath{\mu}{\rm K}}

\newcommand{\msol}{\ensuremath{\mbox{M}_{\odot}}}
\newcommand{\munits}{\times 10^{14}~\msol}
\newcommand{\munitssimple}{10^{14}~\msol}

\newcommand{\lcdm}{{\rm \Lambda CDM}}

\newcommand{\zmin}{0.1}
\newcommand{\zmax}{4.0}
\newcommand{\nlyy}{N_{\ell}^{yy}}
\newcommand{\snrlimit}{5}
\newcommand{\virslope}{A_{\rm v}}
\newcommand{\virintercept}{B_{\rm v}}
\newcommand{\alphay}{\alpha_{_{Y}}}
\newcommand{\betay}{\beta_{_{Y}}}
\newcommand{\gammay}{\gamma_{_{Y}}}
\newcommand{\alphasigmalogy}{\alpha_{\sigma}}
\newcommand{\gammasigmalogy}{\gamma_{\sigma}}
\newcommand{\YszM}{Y_{_{\rm SZ}}-M}
\newcommand{\hsebias}{b_{_{\rm HSE}}}
\newcommand{\hsebiasval}{0.2}

\newcommand{\hseventy}{{\rm h_{70}}}
\newcommand{\thetamax}{2^{\prime}}

\newcommand{\fskywideclean}{50\%}
\newcommand{\fskywidedirty}{17\%}
\newcommand{\fskywidefull}{67\%}
\newcommand{\fskydeepfull}{3\%}
\newcommand{\nth}{_{\rm nth}}

\newcommand{\howmanyclustersforcmbhdbaseline}{325860}
\newcommand{\howmanyclustersforsfourbaseline}{75701}
\newcommand{\howmanyclustersforsfourdeepbaseline}{13699}
\newcommand{\howmanyhighzclustersforcmbhdbaseline}{11095}
\newcommand{\howmanyhighzclustersforsfourbaseline}{992}
\newcommand{\howmanyhighzclustersforsfourdeepbaseline}{341}
\newcommand{\howmanyclustersforcmbhddirty}{76165}
\newcommand{\howmanyclustersforsfourdirty}{17541}
\newcommand{\howmanyhighzclustersforcmbhddirty}{1894}
\newcommand{\howmanyhighzclustersforsfourdirty}{166}

\FPeval{\howmanyclustersforcmbhdfull}{round(\howmanyclustersforcmbhdbaseline + \howmanyclustersforcmbhddirty, 0)}
\FPeval{\howmanyhighzclustersforcmbhdfull}{round(\howmanyhighzclustersforcmbhdbaseline + \howmanyhighzclustersforcmbhddirty, 0)}
\FPeval{\howmanyclustersforsfourfull}{round(\howmanyclustersforsfourbaseline + \howmanyclustersforsfourdirty,0)}
\FPeval{\howmanyhighzclustersforsfourfull}{round(\howmanyhighzclustersforsfourbaseline + \howmanyhighzclustersforsfourdirty,0)}

\newcommand{\medianzcmbhd}{0.7}
\newcommand{\medianzsfour}{0.8}
\newcommand{\medianzsfourdeep}{0.7}

\newcommand{\medianmasscmbhd}{0.8}
\newcommand{\medianmasssfour}{1.6}
\newcommand{\medianmasssfourdeep}{1.0}
\newcommand{\lensingmasserrorcmbhd}{0.002} 
\newcommand{\lensingmasserrorsfour}{0.02} 
\newcommand{\lensingmasserrorsfourdeep}{0.05} 

\newcommand{\medianmasshighzcmbhd}{0.4}
\newcommand{\medianmasshighzsfour}{0.8}
\newcommand{\medianmasshighzsfourdeep}{0.6}
\newcommand{\lensingmasserrorhighzcmbhd}{0.02} 
\newcommand{\lensingmasserrorhighzsfour}{0.31} 
\newcommand{\lensingmasserrorhighzsfourdeep}{0.55} 

\newcommand{\vireffetafid}{1.0}
\newcommand{\vireffetasmall}{0.9}
\newcommand{\vireffetalarge}{1.1}

\newcommand{\virslopeval}{0.155}
\newcommand{\virinterceptval}{0.189}

\newcommand{\shortauthourlist}{Raghunathan, Whitehorn, Alvarez, \it{et al.}}
\newcommand{\authourlist}
{
\author[0000-0003-1405-378X]{Srinivasan Raghunathan}
\affiliation{Center for AstroPhysical Surveys, National Center for Supercomputing Applications, Urbana, IL 61801, USA}
\affiliation{Department of Physics and Astronomy, Michigan State University, 567 Wilson Road, East Lansing, MI 48824}
\affiliation{Department of Physics and Astronomy, University of California, Los Angeles, CA 90095, USA}

\author[0000-0002-3157-0407]{Nathan Whitehorn}
\affiliation{Department of Physics and Astronomy, Michigan State University, 567 Wilson Road, East Lansing, MI 48824}
\affiliation{Department of Physics and Astronomy, University of California, Los Angeles, CA 90095, USA}

\author[0000-0002-2796-9650]{Marcelo A. Alvarez} 
\affiliation{Lawrence Berkeley National Laboratory, One Cyclotron Road, Berkeley, CA 94720, USA}
\affiliation{Berkeley Center for Cosmological Physics, UC Berkeley, CA 94720, USA}

\author[0000-0002-2153-6096]{Han Aung}
\affiliation{Department of Physics, Yale University, New Haven, CT 06520, USA}

\author[0000-0001-5846-0411]{Nicholas Battaglia}
\affiliation{Department of Astronomy, Cornell University, Ithaca, NY 14853, USA}

\author[0000-0002-0463-6394]{Gilbert P. Holder}
\affiliation{Astronomy Department, University of Illinois at Urbana-Champaign, 1002 W. Green Street, Urbana, IL 61801, USA}
\affiliation{Department of Physics, University of Illinois Urbana-Champaign, 1110 W. Green Street, Urbana, IL 61801, USA}

\author[0000-0002-6766-5942]{Daisuke Nagai}
\affiliation{Department of Physics, Yale University, New Haven, CT 06520, USA}

\author[0000-0002-7957-8993]{Elena Pierpaoli}
\affiliation{Physics \& Astronomy Department, University of Southern California, Los Angeles, California, 90089-0484}

\author[0000-0003-2226-9169]{Christian L. Reichardt}
\affiliation{School of Physics, University of Melbourne, Parkville, VIC 3010, Australia}

\author[0000-0001-7192-3871]{Joaquin D. Vieira}
\affiliation{Center for AstroPhysical Surveys, National Center for Supercomputing Applications, Urbana, IL 61801, USA}
\affiliation{Astronomy Department, University of Illinois at Urbana-Champaign, 1002 W. Green Street, Urbana, IL 61801, USA}
\affiliation{Department of Physics, University of Illinois Urbana-Champaign, 1110 W. Green Street, Urbana, IL 61801, USA}

\correspondingauthor{Srinivasan Raghunathan}\email{srinirag@illinois.edu}
}

\newcommand{\tittext}{Constraining Cluster Virialization Mechanism and Cosmology using Thermal-SZ-selected clusters from Future CMB Surveys}

\newcommand{\abstracttext}{
We forecast the number of galaxy clusters that can be detected via the thermal Sunyaev-Zel{'}dovich (tSZ) signals by future cosmic microwave background (CMB) experiments, primarily the wide area survey of the CMB-S4 experiment but also CMB-S4's smaller delensing survey and the proposed \cmbhd{} experiment.
We predict that CMB-S4 will detect 75,000 clusters with its wide survey of $\fsky=$ \fskywideclean{} and 14,000 clusters with its deep survey of $\fsky=$ \fskydeepfull.
Of these, approximately 1350 clusters will be at $z \ge 2$, a regime that is difficult to probe by optical or X-ray surveys.
We assume \cmbhd{} will survey the same sky as the \sfour{}, and find that \cmbhd{} will detect $\times3$ more overall and an order of magnitude more $z \ge 2$ clusters than CMB-S4.
These results include galactic and extragalactic foregrounds along with atmospheric and instrumental noise.
Using CMB-cluster lensing to calibrate cluster tSZ-mass scaling relation, we combine cluster counts with primary CMB to obtain cosmological constraints for a two parameter extension of the standard model ($\lcdm+\summnu+\wde$).
Besides constraining $\sigma(\wde)$ to $\lesssim 1\%$, we find that both surveys can enable a $\sim 2.5-4.5\sigma$ detection of $\summnu$, substantially strengthening CMB-only constraints.
We also study the evolution of intracluster medium by modelling the cluster virialization ${\rm v}(z)$ and find tight constraints from CMB-S4, with further factors of 3--4 improvement for CMB-HD. 
The binned cluster counts, Fisher matrices, and other associated products can be downloaded from this \href{https://github.com/sriniraghunathan/tSZ_cluster_forecasts}{link$^{\text{\faGithub}}$}.
}

\begin{document}

\title{\tittext}
\shorttitle{Cluster Virialization and Cosmological Constraints from CMB Surveys}

\shortauthors{\shortauthourlist}
\authourlist

\begin{abstract}
\abstracttext{}
\end{abstract}


\section{Introduction}
Galaxy clusters are the largest and most massive gravitationally bound systems in the Universe. 
They form on the densest points of the cosmic web and hence contain a wealth of information about structure formation. 
Specifically, cluster abundance as a function of mass and redshift is sensitive to cosmological parameters that govern the geometry and structure growth in the Universe. 
Of further importance is the different degeneracy direction between structure growth parameters probed by clusters compared to cosmic microwave background (CMB) or Baryonic Acoustic Oscillations (BAO) which provides compelling joint constraints.
This has been demonstrated previously in the literature \citep[for example,][]{mantz08, vikhlinin09, rozo10, linden14, dehaan16, salvati17, bocquet19, zubeldia19, planck20_cosmo} and 
the potential of clusters as cosmological probes from future surveys has also been a subject of extensive study \citep[for example, ][and recently \citealt{louis17, madhavacheril17, cromer19, gupta20}]{holder01a, lima04, sartoris12, mak13}. 

Hot electrons in the intracluster medium (ICM) transfer energy to CMB photons through inverse Compton scattering \citep{sunyaev70}. This thermal {\it Sunyaev-Zeldovich} effect (tSZ) has been used to detect clusters from CMB surveys \citep{bleem15, planckSZcat16, hilton18, hilton21, huang20, bleem20} and the number of clusters has been rapidly growing from a few hundreds to thousands with the increase in sensitivity of the CMB surveys \citep{benson14, henderson16, bender18}.
Future CMB surveys like \cmbhd{} \citep{sehgal19}, CMB-S4 \citep{cmbs4collab19}, CORE \citep{melin17} and Simons Observatory (SO, \citealt{SO18}) will increase the sample size by several fold producing mass-limited cluster catalogs down to $\mvir \lessapprox 10^{14}\ \msol$.
In the context of galaxy clusters, future CMB surveys play an important and unique role as they open the window into high redshift $z \gtrsim 2$ Universe using the redshift independent tSZ effect enabling the detection of distant clusters. 
These distant clusters will otherwise be hard to detect using optical or X-ray surveys and as a result future tSZ-selected cluster samples will be complementary to the ones from Large Synoptic Survey Telescope (LSST) at the Vera C. Rubin Observatory \citep{lsst09}, Euclid \citep{laureijs11}, and eROSITA \citep{erosita12}.

The clusters also gravitationally lens the background CMB, an effect known as {\it CMB-cluster lensing}.
After the first set of detections using CMB temperature by Atacama Cosmology Telescope (ACT, \citealt{madhavacheril15}), South Pole Telescope (SPT, \citealt{baxter15}), and \planck{} \citep{melin15, plancksz15},  the field has rapidly evolved to use the signal to calibrate richness-mass scaling relations of optically selected galaxy clusters \citep[][for example]{geach17} and to warrant the first polarization-only detection of the signal by SPTpol survey \citep{raghunathan19c}. 
Like the tSZ effect, CMB-cluster lensing also plays a key role in facilitating the mass measurements of distant clusters expected from the future CMB surveys. 
This is difficult with galaxy weak lensing since the signal-to-noise (\snr) of lensed background galaxies drop significantly at high redshifts. 
\citet{plancksz15} and \citet{zubeldia19} used CMB-cluster lensing information to derive cosmological constraints with \planck{} cluster sample while \citet{alonso16} and \citet{madhavacheril17} studied the potential of CMB-cluster lensing either independently or in combination with galaxy weak-lensing to calibrate the observable-mass (\mbox{$\YszM$}) scaling relations of clusters from CMB-S4 and its impact on cluster cosmology. 
While extensive studies highlighting importance of clusters as cosmological probes exist in the literature, understanding the virialization mechanism and gastrophysics of high redshift ($z \ge 2$) clusters mostly remains an unexplored territory owing to the lack of observations. 

In this work, we focus on 
astrophysical and cosmological constraints using cluster samples from future CMB surveys. 
Our primary focus is on the wide area survey (\sfour) of the CMB-S4 experiment but we also provide predictions for the smaller but deeper CMB-S4 delensing survey (\sfourdeep). 
The proposed \cmbhd{} experiment is also added to the list as an ideal case. 
We start by forecasting the number of tSZ-selected galaxy clusters from the three surveys. 
The simulations used for forecasting are designed to capture most of the effects expected in a real survey. 
They contain atmospheric and instrumental noise along with signals from galactic and astrophysical foregrounds. 
The detected clusters are then binned in lensing mass (obtained using CMB-cluster lensing), tSZ \snr{} $q$, and redshift. 
We combine the binned cluster counts $\binnedcluscounts$ with primary CMB temperature and polarization spectra to derive parameter constraints. 
Besides cosmology and \mbox{$\YszM$} scaling relation, we also study the evolution of the ICM using high redshift clusters. 
For this, we modify the tSZ signals of clusters, using two parameterizations of virialization model ${\rm v}(z) = 1 - \dfrac{Y\nth}{Y_{\rm tot}}$ where $Y_{\rm th}$ and $Y\nth$ are the thermal and non-thermal components of the total integrated Compton-$y$ signal $Y_{\rm tot} = Y_{\rm th} + Y\nth$. 
In the first approach, we use a linear model to scale the tSZ signals from clusters with $z \ge 2$. 
We note that this toy model with a step function at $z \ge 2$ is over simplistic to accurately capture the redshift dependence of the cluster virialization process. 
For example, \citet{fakhouri10} showed that mergers, which are considered as an important source of non-thermal pressure, increase as a function of redshift which would modify the virialization mechanism of high redshift clusters. 
To take this into account, we build a second more realistic model using a fitting formalism, ${\rm v}(z) = \virslope {\rm ln}(1+z) + \virintercept$, that has been derived using the analytic model of the non-thermal pressure in the ICM \citep{Shi14, Green20} and tested using Omega500 simulations \citep{nelson14a, Shi15}.

This paper is structured as follows: We describe the simulation components, cluster virialization model, detection algorithm, mass calibration using CMB lensing and the Fisher formalism to combine binned cluster counts with primary CMB information in \S\ref{sec_methods}. In \S\ref{sec_nlyy_ilc} and \S\ref{sec_baseline_results}, we discuss the baseline results including cluster detection sensitivity, survey completeness, and cluster counts. The modification to cluster sensitivity and counts due to changes in virialization mechanism are given in \S\ref{sec_sensitivity_vir_model_dependence}. We discuss the Fisher forecasts along with the impact of several choices we make in \S\ref{sec_forecasts} and \S\ref{sec_constraints_s4years}. We test the effect of cluster correlated foreground signals in \S\ref{sec_clus_corr_fg} and finally conclude in \S\ref{sec_conclusion}.

Throughout this work, we use \planck{} 2015 cosmology (TT+ lowP in Table 4 of \citealt{planck15-13}) and report cluster masses in units of $\mvir$ which is the mass within a sphere of radius $\rvir$ where the density is 500 times the critical density of the Universe at the cluster redshift. 


\section{Methods}
\label{sec_methods}

\subsection{Simulations}
\label{sec_sims_overview}

The simulations used for this study are \boxsize{} wide CMB temperature realizations with a pixel resolution of $\pixres$. 
Besides primary CMB, the simulations also contain the following frequency dependent signals: cluster tSZ, galactic and astrophysical foregrounds, and experimental noise (both atmospheric and instrumental).
The underlying power spectrum used to generate the primary CMB is the large-scale structure lensed temperature spectrum $C_{\ell}^{TT}$ for the fiducial \planck{} 2015 cosmology \citep{planck15-13}
obtained using \texttt{CAMB} \citep{lewis00} software. 
Cluster tSZ signal is modelled using a generalized Navarro-Frenk-White (NFW, \citealt{navarro96, zhao96, nagai07, arnaud10}) profile as described below in \S\ref{sec_clustertsz}. 
Galactic and astrophysical foreground modellings are in \S\ref{sec_foregrounds}. 
The simulated maps are then convolved by experimental beam functions, assumed to be Gaussian (see Table~\ref{tab_exp_specs}). 
Finally, we add noise realizations to the simulated maps. 
We model the noise spectra to include both atmospheric and instrumental noise as \citep{tegmark97}
\begin{equation}
N_{\ell} = \Delta_{T}^{2} \left[ 1 + \left( \dfrac{\ell}{\ell_{\rm knee}}\right)^{-\alpha_{\rm knee}} \right], 
\end{equation}
where $\Delta_{T}^{2}$ corresponds to detector white noise while $\ell_{\rm knee}$ and $\alpha_{\rm knee}$ are used to model the atmospheric $1/f$ noise. 
This parameterization gives us a sense of the range of multipoles being affected by the atmospheric ($\ell < \ell_{\rm knee}$) and instrumental noise components ($\ell \ge \ell_{\rm knee}$).

\subsection{Experimental setup}
\label{sec_exp_setup}

\begin{deluxetable*}{| l | c | ccccccc | ccccccc |}[h]
\tabletypesize{\footnotesize}
\def\arraystretch{1.2}
\tablecaption{Experimental specifications}
\label{tab_exp_specs}
\tablehead{
\multirow{2}{*}{Experiment} & \multirow{2}{*}{Location ($f_{\rm sky}$)} & \multicolumn{7}{c|}{Beam $\theta_{\rm FWHM}$ [arcminutes]} & \multicolumn{7}{c|}{$\Delta_{T}$ [$\ukam$]}\\
\cline{3-16}
& & 30 & 40 & 90 & 150 & 220 & 270 & 350 & 30 & 40 & 90 & 150 & 220 & 270 & 350
}
\startdata
\cmbhd{} & \multirow{2}{*}{Chile (\fskywidefull)} & 1.4 & 1.05 & 0.45 & 0.25 & 0.20 & 0.15 & 0.12 & 6.5 & 3.4 & 0.73 & 0.79 & 2.0 & 2.7 & 100 \\
\cline{1-1}\cline{3-16}
\sfour{} & & 7.3 & 5.5 & 2.3 & 1.5 & 1.0 & 0.8 & - & 21.8 & 12.4 & 2.0 & 2.0 & 6.9 & 16.7 & - \\\hline\hline
\sfourdeep{} & South Pole (\fskydeepfull) & 8.4 & 5.8 & 2.5 & 1.6 & 1.1 & 1.0 & - & 4.6 & 2.94 & 0.45 & 0.41 & 1.29 & 3.07 & - \\\hline
\hline
\enddata
\end{deluxetable*}

\begin{deluxetable}{| c | c | c | c |}[b]
\def\arraystretch{1.2}
\tablecaption{Atmospheric $1/f$ noise specifications ($\ell_{\rm knee}$, $\alpha_{\rm knee}$).}
\label{tab_exp_atm_noise}
\tablehead{
Band [GHz] & \cmbhd & \sfour & \sfourdeep}
\startdata
30 & \multicolumn{2}{c|}{471, 3.5}& 1200, 4.2\\\hline
40 & \multicolumn{2}{c|}{478, 3.5}& 1200, 4.2\\\hline
90 & \multicolumn{2}{c|}{2154, 3.5}& 1200, 4.2\\\hline
150 & \multicolumn{2}{c|}{4364, 3.5}& 1900, 4.1\\\hline
220 & \multicolumn{2}{c|}{7334, 3.5}& 2100, 3.9\\\hline
270 & \multicolumn{2}{c|}{7308, 3.5}& 2100, 3.9\\\hline
350 & 7500, 3.5 & - & - \\\hline\hline
\enddata
\end{deluxetable}

We consider three future CMB surveys in this work: \cmbhd, \sfour, and \sfourdeep.
Table~\ref{tab_exp_specs} lists the instrumental noise levels $\Delta_{T}\ (\ukam)$ and experimental beams of each frequency band for the three surveys. 
\cmbhd{} is a proposed high resolution millimeter-wave (mm) survey scanning large regions of the sky from Chile with a 30-metre primary mirror and designed to operate in seven bands from 30-350 GHz \citep{sehgal19, sehgal20}. 
CMB-S4 is an upcoming survey that is currently in its design stages and expected to start operations later this decade \citep{cmbs4collab19}. 
In this work, we only consider the two CMB-S4 large aperture telescope (LAT) surveys and not the small aperture telescope survey that is aimed at the inflationary gravitational waves. 
The two S4-LAT surveys (\sfour{} and \sfourdeep) will be performed using 6-metre class telescopes in six\footnote{\sfourdeep{} is also expected to have a 20 GHz band but we ignore that in this work for simplicity.} frequency bands from 30-270 GHz. 
\sfour{} is a legacy survey from Chile and will cover roughly \fskywidefull{} of the sky area. 
\sfourdeep{} is the ``delensing'' survey that is aiming to provide deep observations of $\sim$3\% of the sky from the South Pole. 
The primary objective of \sfourdeep{} survey is to generate high resolution maps of the dark matter distribution in the Universe to facilitate the detection of inflationary $B-$modes by cleaning the lensing induced $B-$modes. 
However, given the large telescope size, \sfourdeep{} also has the capability to detect high redshift SZ clusters as we show in this work.
The parameters governing the atmospheric $1/f$ noise, $\ell_{\rm knee}$ and $\alpha_{\rm knee}$, are listed in Table~\ref{tab_exp_atm_noise} \citep{cmbs4collab19}. 
For simplicity, we assume the $1/f$ model and sky fraction for \cmbhd{} to be the same as the Chile-based \sfour{} survey. 

\subsection{Foreground signals}
\label{sec_foregrounds}
Although extragalactic foregrounds, emissions from dusty star forming galaxies (DSFG) 
in particular, are expected to be the major source of contamination for cluster detection, the footprint of Chile-based experiments cover \fskywidefull{} of the sky area and will be subjected to contamination from galactic emission. 
Hence, we consider both galactic and extragalactic foreground signals for \cmbhd{} and \sfour. 
For \sfourdeep{} we only include extragalactic foregrounds as it will observe a relatively clean region of the sky shown as yellow dashed contours in Fig.~\ref{fig_gal_emission}. 

\subsubsection{Galactic emission}
\label{sec_galactic_foregrounds}

\begin{figure*}
\centering
\ifdefined\ApJsubmit
\includegraphics[width=\textwidth, keepaspectratio]{pysm3_dust_145ghz_high_low_gal_regions.pdf}
\else
\includegraphics[width=\textwidth, keepaspectratio]{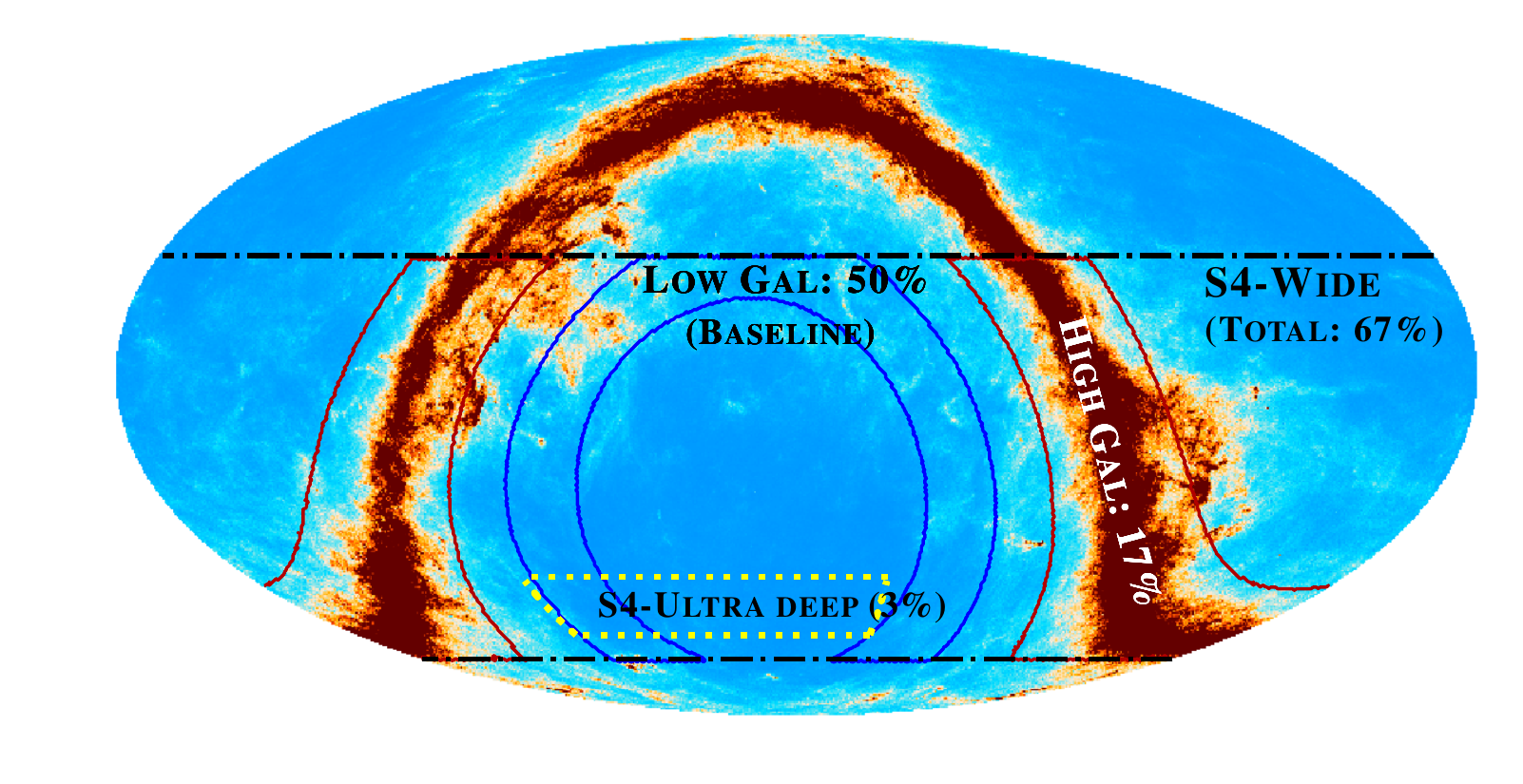}
\fi
\caption{Map of the galactic dust emission at 150 GHz from pySM3 simulations. 
The expected footprints for \sfour{} ($\fsky = \fskywidefull$) and \sfourdeep{} ($\fsky = \fskydeepfull$) are highlighted in black and yellow. 
Regions of high and low galactic emissions used for injecting galactic foregrounds in our simulations are marked as red and blue contours. 
We assume the galactic emission in 23\% of \sfour{} footprint ($\fsky$ = \fskywidedirty) to be similar to high galactic emission in red contour and the emission in the remaining 77\% ($\fsky$ = \fskywideclean) to be similar to low galactic emission in blue contour (marked as baseline). 
This is a conservative choice as the blue contour is not the cleanest region in the \sfour{} footprint. 
We use the same \sfour{} footprint and strategy for \cmbhd.
Since \sfourdeep{} will observe in a relatively clean patch, we do not include galactic foregrounds for \sfourdeep{} simulations.
}
\label{fig_gal_emission}
\end{figure*}

The galactic foregrounds signals, dust and synchrotron, are position dependent and hence one cannot rely on Gaussian realizations of an underlying power spectrum for the entire footprint. 
To this end, we use the publicly available pySM3 dust and synchrotron map simulations\footnote{\url{https://github.com/CMB-S4/s4mapbasedsims/tree/master/202102_design_tool_input}.} being built specifically in the context of CMB-S4. 
For more details about pySM3 simulations, we refer the reader to the original work \citep{thorne17} which is partly based on the \planck{} Sky Model code \citep{delabrouille13}. 
We use \texttt{S0\_d0} dust and \texttt{S0\_s0} synchrotron models of pySM where the dust temperature, dust emissivity index, and synchrotron spectral index does not have spatial variations. 
The models also ignore any non-Gaussianities. 
Other galactic signals like free-free and anomalous microwave emissions, which should have negligible impact on cluster searches, are ignored in this work. 
To estimate the position dependent galactic foregrounds, we first divide the \sfour{} footprint into two: {\it High} and {\it low} emission regions, shown as red and blue contours in Fig.~\ref{fig_gal_emission}.
High emission region corresponds to $\pm 15^{\circ}$ and encompasses most of signals from the galactic plane. 
Low emission region corresponds to regions in the range $-45^{\circ} \le b \le -30^{\circ}$. 
We compute the temperature power spectra $C_{\ell}^{\rm gal}$ of dust and synchrotron signals of both these regions for all the frequency bands of interest. 

We assume that 23\% ($\fsky$ = \fskywidedirty) of \sfour{} footprint to have galactic signals similar to high emission region and the galactic signals in the remaining 77\% ($\fsky$ = \fskywideclean) to be similar to low emission region. 
As evident from the figure, the blue contour does not correspond to the cleanest region in \sfour{} footprint and hence this is a conservative choice. 
With this assumption, we use the $C_{\ell}^{\rm gal}$ spectra estimated in the two regions to generate Gaussian realizations of galactic emission and add them to our maps to produce two sets of simulated skies. 
The underlying power spectrum for the other signals in the two sets are the same. 
By doing this, we approximate the galactic power spectra to be constant inside the two regions. 
We validated this assumption by dividing the \sfour{} footprint into six regions (latitude steps of $\Delta b = 15^{\circ}$) with different galactic emission and do not find significant difference in the number of detected clusters between the two approaches. 
We follow the same approach for \cmbhd. 
Like mentioned above, since the footprint of \sfourdeep{} lies in a relatively clean region, we do not include galactic foregrounds for \sfourdeep{} simulations. 

The CMB-S4 pySM3 simulations does not include the 350~GHz band that is required for \cmbhd. 
We obtain auto spectra of the galactic dust at 350 GHz band and its cross correlation with other bands by simply scaling the 270 GHz dust spectra as   
\begin{equation}
\label{eq_gal_dust}
C_{\ell, \nu_1 \nu_2}^{\rm gal-dust} = C_{\ell, \nu_{0} \nu_{0}}\ \epsilon_{\nu_{1}, \nu_{2}}\ \dfrac{\eta_{\nu_{1}} \eta_{\nu_{2}}}{\eta_{\nu_{0}} \eta_{\nu_{0}}}, 
\end{equation}
where $\nu_{0}=$ 270 GHz and $\nu_1$, $\nu_2$ $\in$ [30, 40, 90, 150, 220, 270, 353] GHz. The terms $\epsilon_{\nu}$ and $\eta_{\nu}$ in Eq.(\ref{eq_gal_dust}) when combined represent the spectral energy distribution ($f_{\nu}$, SED) of dust and we use a modified blackbody of the form
\begin{eqnarray}
\label{eq_dust_freq_dep}
\eta_{\nu} = \nu^{\beta_{d}}\ B_{\nu}(T_{d}),
\end{eqnarray}
and 
\begin{eqnarray}
\label{eq_dust_epsilon}
\epsilon_{\nu_{1}, \nu_{2}} =  \left. \frac{\dfrac{d B_{\nu_{0}}}{dT} \dfrac{d B_{\nu_{0}}}{dT}} {\dfrac{d B_{\nu_{1}}}{dT}\dfrac{d B_{\nu_{2}}}{dT}} \right|_{T = T_{\rm CMB}}.
\end{eqnarray}
$B_{\nu}(T)$ in the above equations is the \planck{} function and we set $T_{\rm CMB}$ = 2.73 K, emissivity index $\beta_{d} = 1.6$, and dust temperature $T_{d}$ = 19.6 K similar to pySM3 simulations that are consistent with measurements from \planck{} \citep{planck18XI}. 
We ignore synchrotron signals at 350 GHz since they are expected be negligible compared to dust at such high frequencies.

\subsubsection{Extragalactic foregrounds}
\label{sec_exgal_foregrounds}
Diffuse extragalactic foreground signal can be decomposed into: emissions from DSFG and radio galaxies (RG) below the detection threshold; and kinetic SZ (kSZ) and tSZ signals. 
Note that DSFGs are responsible for cosmic microwave background (CIB) anisotropies at mm/sub-mm wavelengths and we sometimes use the two terms, DSFG and CIB, interchangeably in this work. 
DSFG and RG signals were all modelled as Gaussian realizations using SPT power spectra measurements \citep{george15, reichardt21}. 
SPT observations masked DSFG and RG detected above $5\sigma$ which corresponds to a flux threshold of $S_{150} \sim 6$ mJy. 
However, the $5\sigma$ detection limit for the future surveys considered here will be much lower: $S_{150} \sim$2 mJy for CMB-S4 \citep{cmbs4collab19} and $\le 0.1$ mJy for \cmbhd{} \citep{sehgal19}. 
For CMB-S4, we do not modify the masking threshold and simply use SPT measurements. 
Thus, DSFG and RG signals injected into \sfour{} and \sfourdeep{} simulations are conservative estimates. 
For \cmbhd{}, however, \citet{sehgal19} claim that sources with flux above $S_{150} \ge$ 0.04 mJy can be efficiently removed by detecting them at $\ge 3\sigma$ in 270/350 GHz bands. 
This lowers the DSFG power in 150 GHz by $\times17$ and we adopt this strategy here. 
The masking threshold for RG is not modified from SPT values. 
The frequency dependence of DSFG and RG signals are also adopted from SPT \citep{george15, reichardt21} measurements. 
We introduce decorrelation in the DSFG signals between 270/350 GHz and 150 GHz bands using SPT $\times$ {\it Herschel}/SPIRE measurements \citep{viero19}. 
We estimate the correlation coefficient between 150 GHz and 270/350 GHz bands by interpolating the values in Table 1 of \citet{viero19}. 

The kSZ signal is contributed by two distinct sources: one from the Doppler boosting of CMB photons due to the motion of haloes and the other from the epoch of reionization. 
We use \citet{reichardt21} measurement to simulate the kSZ signal. 
This roughly corresponds to flat spectrum in $D_{\ell}$ with $D_{\ell, 3000} = 3\ \uk^{2}$ where $D_{\ell} = C_{\ell}\ \dfrac{\ell(\ell+1)}{2\pi}$ and has no frequency dependence. 
For the diffuse tSZ, we consider power from all haloes with $\mvir \ge 10^{13}~\msol$, modelled using Arnaud profile \citep{arnaud10}, in the redshift range $z \in [\zmin, \zmax]$. 
Both the diffuse kSZ and tSZ signals are simulated as Gaussian realizations using the respective power spectrum described above. 

In our fiducial setup, DSFG/RG/kSZ signals are assumed to be uncorrelated to the cluster under study. 
This is, however, not entirely correct as galaxies preferentially reside inside clusters and the cluster motion can also give rise to kSZ signals. We test this assumption in \S\ref{sec_clus_corr_fg} by injecting cluster correlated foreground signals using Websky \citep{stein20} and \mbox{MultiDark Planck 2} (MDPL2, \citealt{omori21prep}) simulations. 

\subsection{Cluster tSZ signal}
\label{sec_clustertsz}

We model the ICM pressure using the dimensionless universal pressure profile $P_{e}(x)$ proposed by \citet{nagai07} and calibrated using X-ray observations by \citet{arnaud10}
\begin{equation}
\label{eq_gnfw_arnaud_profile}
P_{e}(l, x) = \dfrac
{ P_{0} }
{ 
(c_{500}x)^{\gamma} \left[1 + (c_{500}x)^{\alpha}\right]^{\left(\frac{\beta-\gamma}{\alpha} \right)}
},
\end{equation}
where the distance to the cluster centre $x\equiv xR_{500}$, expressed in terms of virial radius $R_{500}$, and concentration parameter $c_{500}$ are related to scale radius $r_{s}$ as $x = r/r_{s}$ and $c_{500} = R_{500}/r_{s}$. 
The best-fit values of parameters are \citep{arnaud10}: 
$c_{500}= 1.177$, the normalization constant $P_{0} = 8.403\ \hseventy^{-3/2}$, and the exponents are $\alpha = 1.0510$, $\beta = 5.4905$, and $\gamma = 0.3081$. 
The pressure profile $P_{e}(l, x)$ is integrated along the line-of-sight to obtain the Compton-$y$ signal $y(x)$ as
\begin{equation}
\label{eq_compton_y}
y(x) =  \frac{\sigma_{T}}{m_{e} c^{2}} \int_{l}P_{e}(l, x)\ dl
\end{equation} 
and converted into CMB temperature units as
\begin{equation}
\label{eq_compton_y_to_tsz}
\delta_{T} = y(x)\ g_{\rm SZ}(\nu) T_{\rm CMB}\ {\rm K},
\end{equation} 
where $\sigma_{T}$ in the Thomson cross section, $c$ in the velocity of light, $m_{e}$ is the electron mass, $T_{\rm CMB} = 2.73\ {\rm K}$ is the mean temperature of the CMB and $g_{\rm SZ}(\nu)$ is the frequency dependence of the tSZ signal which, ignoring relativistic SZ corrections \citep[e.g.,][]{itoh98,chluba12}, is given by 
\begin{eqnarray}
\label{eq_tsz_freq_dep}
g_{\rm SZ}(\nu)  & = & x\ {\rm coth}(x/2) - 4;\ x = \frac{h \nu}{k_{B}T_{\rm CMB}}, 
\end{eqnarray} where $h$ and $k_{B}$ are Planck and Boltzmann constants respectively. 

We integrate $y(x)$ over the angular extent of the cluster $R_{500}$ to obtain the total integrated cluster Compton $\Yvir$ defined using \citet{plancksz15} but generalized based on \citet{alonso16} and \citet{madhavacheril17} to include mass and redshift evolution as
\begin{eqnarray}
\label{eq_Ysz_mass}
\Yvir = {\rm v(z)}\ Y_{\ast}\ \left[\dfrac{h}{0.7}\right]^{-2+\alphay}\ \left[ \dfrac{\mvir}{M_{\ast}} \right]^{\alphay} \\\notag
{\rm e}^{\betay {\rm log}^{2} \left( \frac{\mvir}{M_{\ast}} \right)}\ \left[\dfrac{D_{A}(z)}{100 {\rm Mpc}}\right]^{2} E^{2/3}(z)\  (1+z)^{\gammay},
\end{eqnarray}
where $M_{\ast} = 6 \munits$ is the pivotal mass, $D_{A}(z)$ is the angular diameter distance to the cluster at redshift $z$, $E(z) = H(z)/H_{0}$ is the Hubble function and $\mvir$ is the mass of the cluster. 
${\rm v}(z)$ in the above equation is the cluster virialization model adopted to modify the cluster tSZ signal and is explained below in \S\ref{sec_vir_model}.
We use \citet{plancksz15} best-fit values to fix ${\rm log}Y_{\ast} = -0.19$ and $\alphay = 1.79$. 
The fiducial values for redshift $\gammay$ and second-order mass $\betay$ evolution parameters are set to zero. 
The log-normal scatter $\sigma_{{\rm log}\Yvir} \equiv \sigma_{{\rm log}Y}$ in the above relation is modelled similar to \citet{madhavacheril17} to include mass and redshift evolution as 

\begin{eqnarray}
\label{eq_Ysz_mass_scatter}
\sigma_{{\rm log}Y}(\mvir, z) = \sigma_{{\rm log}Y, 0} \left[ \dfrac{ \mvir}{M_{\ast}} \right]^{\alphasigmalogy} (1 + z)^{\gammasigmalogy}
\end{eqnarray}
with the fiducial values set to $\sigma_{{\rm log}Y, 0} = 0.127$, $\alphasigmalogy = 0$, and $\gammasigmalogy = 0$ \citep{louis17}.

\subsection{Modelling the cluster virialization}
\label{sec_vir_model}
Not much is known about the gastrophysics of high redshift clusters owing to the lack of sufficient observations. Lately, \citet{mantz14} and \citet{mantz18} used CARMA data to perform detailed tSZ study of a distant cluster at $z = 1.99_{-0.21}^{+0.19}$ with $\mvir \sim 1-2 \munits$ that was detected by X-ray XMM-{\it Newton} satellite. 
\citet{mantz18} report that the properties of this distant cluster is in reasonable agreement with the extrapolated scaling relations confirming self-similar evolution of clusters out to $z \sim 2$. 
However, the authors also caution the readers about generalizing the result from a single $z\sim2$ cluster to all high redshift clusters. 
As we will see later in \S\ref{sec_baseline_results}, \cmbhd{} and CMB-S4 have the capability to detect hundreds to thousands of clusters with $\mvir \lesssim 10^{14}\ \msol$ at $z \gtrsim 2$. 
Subsequently, we aim to study the physics of the ICM and its evolution out to high redshifts with these potential detections.  
To this end, we tweak the first term in Eq.(\ref{eq_Ysz_mass}), ${\rm v}(z) \equiv \left[1 - \dfrac{Y_{\rm non-th}}{Y_{\rm tot}}\right]$ that controls cluster virialization and as a result modifies the cluster tSZ signal.
We model ${\rm v}(z)$ in two different ways as described below.

\subsubsection{Linear scaling: Model 1}
In the first approach, we use a simple model
\begin{eqnarray}
\label{eq_vir_model_1}
{\rm v}(z) = \vireffeta(z)\ (1- \hsebias)^{\alphay}, 
\end{eqnarray}
where $\hsebias$ is the hydrostatic equilibrium (HSE) mass bias set to $\hsebias = \hsebiasval$ \citep{zubeldia19, ryu20} and assumed to be constant for clusters at all redshifts. $\vireffeta(z)$ is the {\it virialization efficiency} of clusters that modifies the tSZ signal of clusters using a linear scaling as
\begin{eqnarray}
\label{eq_eta_z}
\vireffeta(z) &=&   \left\{
\begin{array}{l l}
1, z < 2\\
1 + \epsilon, z \ge 2
\end{array}\right.
\end{eqnarray}
with $\epsilon \in [-1, 1]$. 
This model is similar to the \mbox{$\YszM$} relation used in \citet{plancksz15} except for the introduction of $\vireffeta(z)$ for high-$z$ clusters. 
The fiducial value of $\vireffeta(z) = 1$ for all clusters. 

\subsubsection{Physically motivated model 2}
Since the step function at $z \ge 2$ used in the above model is highly simplistic, we now build a realistic model to parameterize the redshift dependence of the virialization process \citep[e.g.,][]{fakhouri10} more accurately.
In this second approach, we use the analytic model for modelling the evolution of the non-thermal pressure fraction through the cluster assembly and virialization processes \citep{Shi15} and their impact on the $\YszM$ relation of high redshift clusters using the model presented in \citet{Green20}. 
We summarize the modelling and results in Appendix~\ref{sec_cluster_vir_model2} following which we use a fitting formalism
\begin{eqnarray}
\label{eq_vir_model_2}
{\rm v}(z) = \virslope {\rm ln}(1+z) + \virintercept
\end{eqnarray}
derived from the analytical model of non-thermal pressure \citep{Shi14,Green20} and tested using the Omega500 hydrodynamical cosmological simulation \citep{nelson14a, Shi15}.
We set the fiducial values of the parameters to be $\virslope = \virslopeval$ and $\virintercept = \virinterceptval$.

\subsection{Cluster detection}
\label{sec_cluster_detection}

We combine the simulated maps in different frequency channels $N_{\rm ch}$ optimally using an internal linear combination (ILC) algorithm and create a Compton-$y$ map as
\begin{eqnarray}
y_{\ell} = \sum_{i=1}^{\rm N_{ch}} w_{\ell}^{i} M_{\ell}^{i},
\end{eqnarray}
where the multipole dependent weights $w_{\ell}$ for each frequency channel are computed using the \texttt{SMICA} (Spectral Matching Independent Component Analysis) algorithm \citep{cardoso08, remazeilles11, planck14_smica} as 
\begin{eqnarray}
\newcommand{\clinv}{{\bf C}_{\rm \ell}^{-1}}
w_{\ell} = \frac{\clinv {\bf a}}{{\bf a}^{T} \clinv {\bf a}}.
\label{eq_ilc_weights}
\end{eqnarray}
The matrix ${\bf C}_{\ell}$ has a dimension $\rm N_{ch} \times \rm N_{ch}$ and contains the covariance between simulated maps in multiple frequencies at a given multipole $\ell$. 
The frequency response vector \mbox{${\bf a} = [-5.33, -5.23, -4.36, -2.61, 0.09, 2.27, 5.95]$} contains the tSZ spectrum given in Eq.(\ref{eq_tsz_freq_dep}) for [30, 40, 90, 150, 220, 270, 350] GHz channels.  
The weights in Eq.(\ref{eq_ilc_weights}) for each band are chosen optimally to produce a minimum variance Compton-$y$ map by jointly minimizing the contamination from noise and foreground signals that are uncorrelated with the cluster. 
We do not explicitly null any foreground components using a constrained ILC technique \citep{remazeilles11} but study the effect of cluster correlated foreground signals in \S\ref{sec_clus_corr_fg}.

\subsubsection{Maximum likelihood approach}
\label{sec_ml_fitting}

\newcommand{\Ysymbol}{y}
\newcommand{\recoveredYsymbol}{\hat{\Ysymbol}}
\newcommand{\Ysymbolbold}{\boldsymbol{\Ysymbol}}
\newcommand{\recoveredYsymbolbold}{\boldsymbol{\recoveredYsymbol}}

The resultant ILC Compton-$y$ map is then used to compute the $\snr$ of the cluster tSZ signal using a maximum likelihood based approach. 
For blind cluster searches, however, adopting a multi-band  matched-filtering technique \citep{melin06} would be computationally more feasible as done traditionally in cluster finding using CMB surveys \citep[for example]{bleem15}. 
The two approaches are equivalent. 
Besides the cluster tSZ signal, this map includes variance from diffuse tSZ and also the residual CMB, foreground signals, and noise. 
Using this \boxsize{} ILC $y$ map, we calculate
\begin{equation}
    \label{eq_likelihood}
    -2 \ln{} \mathcal{L} = \sum_{ij} \left(\recoveredYsymbolbold_i - \Ysymbolbold^\mathrm{th}_i \right) \mathbf{\hat{C}}_{ij}^{-1} \left(\recoveredYsymbolbold_j - \Ysymbolbold^\mathrm{th}_j \right)\,,
\end{equation}
where $\Ysymbolbold_{i} \equiv \Ysymbolbold_{i}(\theta)$ is the  azimuthally-averaged profile of the Compton-$y$ signal in bins $i$ of \mbox{$\Delta \theta = \pixres$}, out to a maximum $\theta_\mathrm{max}=\thetamax$.
The chosen $\theta_\mathrm{max}$ encompasses the tSZ signal from majority of the clusters at all redshifts and hence maximizes the \snr. 
Specifically, $\theta_\mathrm{max} \ge \thetavir$ for clusters with $z \gtrsim 0.5$ where $\thetavir = \rvir/D_{A}(z)$ and ${D_{A}(z)}$. 
The measured cluster Compton-$y$ from a cluster with a given mass and redshift is $\recoveredYsymbolbold$. 
We compute theory models $\Ysymbolbold^\mathrm{th}$ for different masses at the cluster redshift using Eq.(\ref{eq_Ysz_mass}) and fit them to the measured $\recoveredYsymbolbold$ signal. 
The covariance matrix $\mathbf{\hat{C}}$ includes contribution from other sources of variance described above. 
It is computed using \mbox{$N = 2500$} simulations as
\begin{eqnarray}
    \label{eq_cov_matrix}
    \mathbf{\hat{C}} = \frac{1}{N-1}\sum\limits_{n = 1}^{N} \left(\recoveredYsymbolbold_i - \left< \recoveredYsymbolbold \right> \right) \left(\recoveredYsymbolbold_i - \left< \recoveredYsymbolbold\right> \right)^{T}\,.
\end{eqnarray}

We use the distribution of best-fits recovered from 100 simulations and compute the \snr{} as the inverse of the $1\sigma$ uncertainty defined by the 16 to 84 percent confidence range. 
For each survey, we compute a $\snr$ look-up table in this way for different clusters in a ($\mvir, z$) grid: \mbox{${\rm log}\mvir \in [13,15.4]\ \msol$} with \mbox{${\rm log}\Delta M = 0.1\ \msol$} and $z \in [0.1, 3]$ with $\Delta z = 0.1$. 
This $\snr$ look-up table is used to select clusters above the detection threshold $\snr \equiv q = \snrlimit$ in later sections.

\subsection{Mass calibration using CMB lensing}
\label{sec_cmb_lensing_mass}
We perform internal mass calibration of clusters using their gravitational lensing signatures on both CMB temperature and polarization anisotropies. 
Cluster kSZ and tSZ signals are expected to introduce significant bias to temperature based lensing reconstruction \citep{raghunathan17a}. 
We mitigate them by employing inpainted-gradient \citep{raghunathan19b} quadratic temperature lensing estimator \citep[QE,][]{hu07}. 
This estimator reconstructs lensing using the lensing-induced correlations between a large-scale and a small-scale temperature anisotropies map. 
Cluster SZ signals, besides lensing, can also introduce such correlations which tend to bias best-fit lensing masses. 
In the inpainted-gradient QE, we remove SZ signals in the large-scale gradient map by estimating the pixel values at the cluster location using information from adjacent pixels. 
For polarization, we use the optimal maximum likelihood estimator (MLE) \citep{raghunathan17a} which reconstructs cluster masses using the lensing-induced changes to pixel-pixel covariance matrix. 
We ignore the covariance between temperature and polarization but note that can slightly degrade the lensing \snr. 

\subsection{Fisher formalism}
\label{sec_fisher_formalism}
We use the Fisher matrix formalism \citep{holder01a} and compute 
\begin{equation}
    \label{eq_fisher}
    F_{\theta_{i}\theta_{j}} = \sum_{\binnedcluscountsquantities} \frac{\partial \binnedcluscounts}{\partial \theta_{i}}\frac{\partial \binnedcluscounts}{\partial \theta_{j}} \frac{1}{\binnedcluscounts}, 
\end{equation}
where $\theta_{i}, \theta_{j}$ are the astrophysical or cosmological parameters to be constrained; 
$\binnedcluscounts$ is the number of clusters in a given lensing mass $M_{L}$, tSZ \snr{} $q$, and redshift $z$ bin; and $1/\binnedcluscounts$ gives the Poisson error in each bin. 
The summation indices $lqz$ run over $M_{L}$, $q$, and $z$ bins described below in \S\ref{sec_binning}.
Cluster number counts in a given bin $\binnedcluscountswithsubscripts \equiv \binnedcluscounts$ can be calculated as 
\begin{widetext}
	\begin{align}
	    \label{eq_cluster_counts}	 
		    \binnedcluscountswithsubscripts = 
		    \int_{z_{i}}^{z_{i+1}}
		    dz 
		   \int_{M_{L_{j}}}^{M_{L_{j+1}}}
		    dM_{L}
		    \int_{q_{k}}^{q_{k+1}}
		    dq \int_{0}^{\infty} dM d\Ysz\ \frac{dV}{dz_{\rm true} d\Omega}\ n(M,z_{\rm true})\ \mathcal{N}\left( z |z_{\rm true}, \sigma_{\rm z} \right)\ \mathcal{N}\left( M_{L} |M, \sigma_{M_{L}} \right) \\\notag
		    \mathcal{N}\left( q | \Ysz/\sigma_{\Ysz}, 1\right)\ \mathcal{N}\left( {\rm log}\Ysz | M, z_{\rm true}, \sigma_{{\rm log}_{\Ysz}} \right), 
    \end{align}
\end{widetext} 
where $n(\mvir,z_{\rm true}) \equiv n(M,z_{\rm true})$ is the \citet{tinker08} halo mass function (HMF), and $\frac{dV}{dz_{\rm true} d\Omega}$ is the volume element.  
We parameterize the probability density functions (PDF) of redshift $z$, lensing mass $M_{L}$, and tSZ \snr{} $q$ using normal distributions $\mathcal{N}(\mu, \sigma)$ with mean $\mu$ and width $\sigma$. 
The scatter in the observable-mass scaling relation $\Ysz-M$ is parameterized using a log-normal $\mathcal{N}\left( {\rm log}\Ysz | M, z_{\rm true}, \sigma_{{\rm log}_{\Ysz}} \right)$ with width $\sigma_{{\rm log}_{\Ysz}}$.
Since the photometric redshift errors for clusters from future surveys are expected to be small compared to the width of the redshift bins described below \citep{lsst09} we neglect redshift errors by setting $z = z_{\rm true}$ with $\sigma_{z} = 0$, (i.e:) we assume $\mathcal{N}\left( z |z_{\rm true}, \sigma_{\rm z} \right)$ to be a $\delta$ function. 
Errors in tSZ flux $\sigma_{\Ysz}$ are obtained using the MLE approach described above and the errors in lensing mass $\sigma_{M_{L}}$ are determined using CMB temperature and polarization based reconstructions. 

\subsubsection{Monte Carlo sampling}
\label{sec_monte_carlo_sampling}
We solve the above integral using a Monte Carlo (MC) sampling approach to estimate $\binnedcluscounts$ and its derivatives $\partial \binnedcluscounts/\partial \theta$ as a function of parameter under consideration $\theta$. 
We start by getting the number of haloes $n(\mvir,z)$ in the following mass and redshift bins using the \citet{tinker08} HMF: $\mvir \in  [10^{13}, 10^{16}]\ \msol$ with $\Delta {\rm} \mvir = 10^{12}\ \msol$ and $0.1 \le z \le 3$ with $\Delta z = 0.1$. 
While statistical uncertainties in the HMF parameters could be potentially important \citep{artis21}, we defer their impact on the results to a future work. 
For each halo, we assign a tSZ flux $\Ysz$ from $\mathcal{N}\left( {\rm log}\Ysz | M, z, \sigma_{{\rm log}_{\Ysz}} \right)$ and an associated tSZ \snr{} $q$ from $\mathcal{N}\left( q | \Ysz/\sigma_{\Ysz}, 1\right)$. 
The tSZ $\snr$ for the halo is obtained by interpolating the $\snr$ look-up table in \S\ref{sec_ml_fitting}. 
Lensing mass and redshifts are also assigned using the distributions $\mathcal{N}(M_{L} |M, \sigma_{M_{L}})$ and $\mathcal{N}\left( z |z_{\rm true}, \sigma_{\rm z} \right)$. 
Next we bin the haloes in lensing mass, \snr, and redshift to obtain binned cluster counts $\binnedcluscounts$ as described below. 
We repeat the MC sampling approach 100 times to ensure the convergence of cluster counts $\binnedcluscounts$. 

\subsubsection{Binning scheme}
\label{sec_binning}
We choose 40 and 25 logarithmic bins for lensing mass and tSZ \snr: $M_{L} \in  [10^{12}, 10^{16}]\ \msol$ and $q \in[5, 500]$.
For redshift we consider four different binning schemes. 
In the baseline case, we use $\Delta z = 0.1$ for $0.1 \le z < 1.5$ and conservatively group all high redshift clusters $1.5 \le z \le 3 $ in one massive redshift bin similar to \citet{madhavacheril17}. 
This is due to the difficulties that will be encountered in measuring redshifts of distant clusters. 
While dedicated follow-up observations are needed to obtain redshifts for clusters at $z \ge 1.5$, the absence of an associated signal in multiple LSST bands will still allow us set a lower limit on the cluster redshifts and we set this threshold to be $z = 1.5$. 
Redshifts of clusters with $z < 1.5$ can be obtained using upcoming optical and X-ray surveys \citep{lsst09, erosita12}. 
We also explore other choices for redshift binning: (i)  an extremely optimistic case of $\Delta z = 0.1$ for all clusters; (ii) a less conservative choice of $\Delta z = 0.1$ for $0.1 \le z < 2$ and $\Delta z = 1$ for $2 \le z \le 3$; and (iii) a pessimistic setting by ignoring clusters at $z > 1.5$.

\subsubsection{Derivatives $\partial \binnedcluscounts/\partial \theta$} 
We estimate derivatives of binned cluster counts $\partial \binnedcluscounts/\partial \theta$ as a function of parameter $\theta$ using a finite difference method. 
For this, the MC sampling approach must be repeated twice for every parameter perturbing $\theta \rightarrow \theta \pm \epsilon_{\theta}$. 
The randomness in sampling, however, can lead to unstable derivatives and we avoid this by only estimating counts $N(M_{L}, q, z)$ at the fiducial values of the parameters. 
For derivatives, we assign weights to haloes based on the ratio of the PDF at the sampled point before and after modifying the parameter values. 
Subsequently, the weights are decomposed into $w_{\rm Poi}$, $w_{\rm tSZ}$, $w_{q}$, $w_{M_{L}}$ and $w_{z}$ with the final weight being the product of all the individual ones.
Here, $w_{\rm Poi} = \dfrac{[n(M,z)]_{\theta \pm \epsilon}}{[n(M,z)]_{\theta}}$ quantifies the change in number of haloes when parameter $\theta$ is modified. 
The other weights ($w_{\rm SZ}$, $w_{q}$, $w_{M_{L}}$ and $w_{z}$) are simply the ratio of respective individual PDFs at the sampled point $\dfrac{ [ \mathcal{N}(\mu, \sigma) ]_{\theta  \pm \epsilon}} { [ \mathcal{N}(\mu, \sigma) ]_{\theta}}$ described in \S\ref{sec_fisher_formalism} and $\theta$ is one of the 16 parameters being constrained: observable-mass scaling relation parameters $[\alphay, \betay, \gammay]$ in Eq.(\ref{eq_Ysz_mass}) and $[\sigma_{{\rm log}Y}, \alphasigmalogy, \gammasigmalogy]$ in Eq.(\ref{eq_Ysz_mass_scatter}); cosmological parameters $[A_{s},\ h,\ \summnu,\ n_{s},\ \omchsq,\ \ombhsq,\ \taure,\ \wde]$; and 
cluster virialization parameters from one of the two models namely $[\vireffeta, \hsebias]$ in Eq.(\ref{eq_vir_model_1}) or $[\virslope, \virintercept]$ in Eq.(\ref{eq_vir_model_2}). 

\subsubsection{CMB TT/EE/TE Fisher matrix}
\label{sec_cmb_fisher}
Along with cluster counts, we also make use of the information from primary CMB temperature and polarization power spectra. 
Since clusters can lens the background CMB, cluster counts will have a non-zero covariance with CMB lensing power spectrum. 
We make a conservative choice and fully ignore information from CMB lensing power spectrum in this work. 
We use lensed CMB spectra but do not correct for the lensing induced correlations. 
Because of the non-zero covariance between clusters and CMB-lensing, we note that this can underestimate the error bars \citep{green17}. 
However, the effect is small and hence we do not consider it. 
In a similar vein, we ignore information from tSZ power spectrum since that must be highly correlated with cluster counts.

We compute CMB Fisher matrices using TT, EE, and TE power spectra ($\ell_{\rm max} = 5000$) obtained using \texttt{CAMB} \citep{lewis00} software for the fiducial \planck{} 2015 cosmology described in \S\ref{sec_sims_overview}. 
The CMB TT, EE, and TE information come from the same experiment under consideration. Although we could include \planck{} information on large-scales and in the regions not covered by the experiments in this work, we avoid them in the baseline setup.
We also avoid adding \sfour{} information in the regions not covered by \sfourdeep.
Like in the case of Compton-$y$ maps, we optimally combine information from all frequency channels using the ILC algorithm to compute the residual noise (see Table~\ref{tab_exp_specs} and Table~\ref{tab_exp_atm_noise}) and foreground spectra (see \S\ref{sec_foregrounds}) in the CMB maps for all the three surveys. 
To generate polarized foregrounds, we assume 2\% (3\%) polarization fractions for dusty star forming (radio) galaxies consistent with measurements from ACT \citep{datta18} and SPT \citep{gupta19}.
Diffuse kSZ and tSZ signals are assumed to be unpolarized.
Information about polarized galactic dust and synchrotron signals come from pySM3 simulations (see \S\ref{sec_galactic_foregrounds}). 


\section{Results and discussion}
\label{sec_results}

\subsection{Noise level in Compton-$y$ map}
\label{sec_nlyy_ilc}

\begin{figure}
\centering
\ifdefined\ApJsubmit
\includegraphics[width=.48\textwidth, keepaspectratio]{compton_nl_yy.pdf}
\else
\includegraphics[width=.48\textwidth, keepaspectratio]{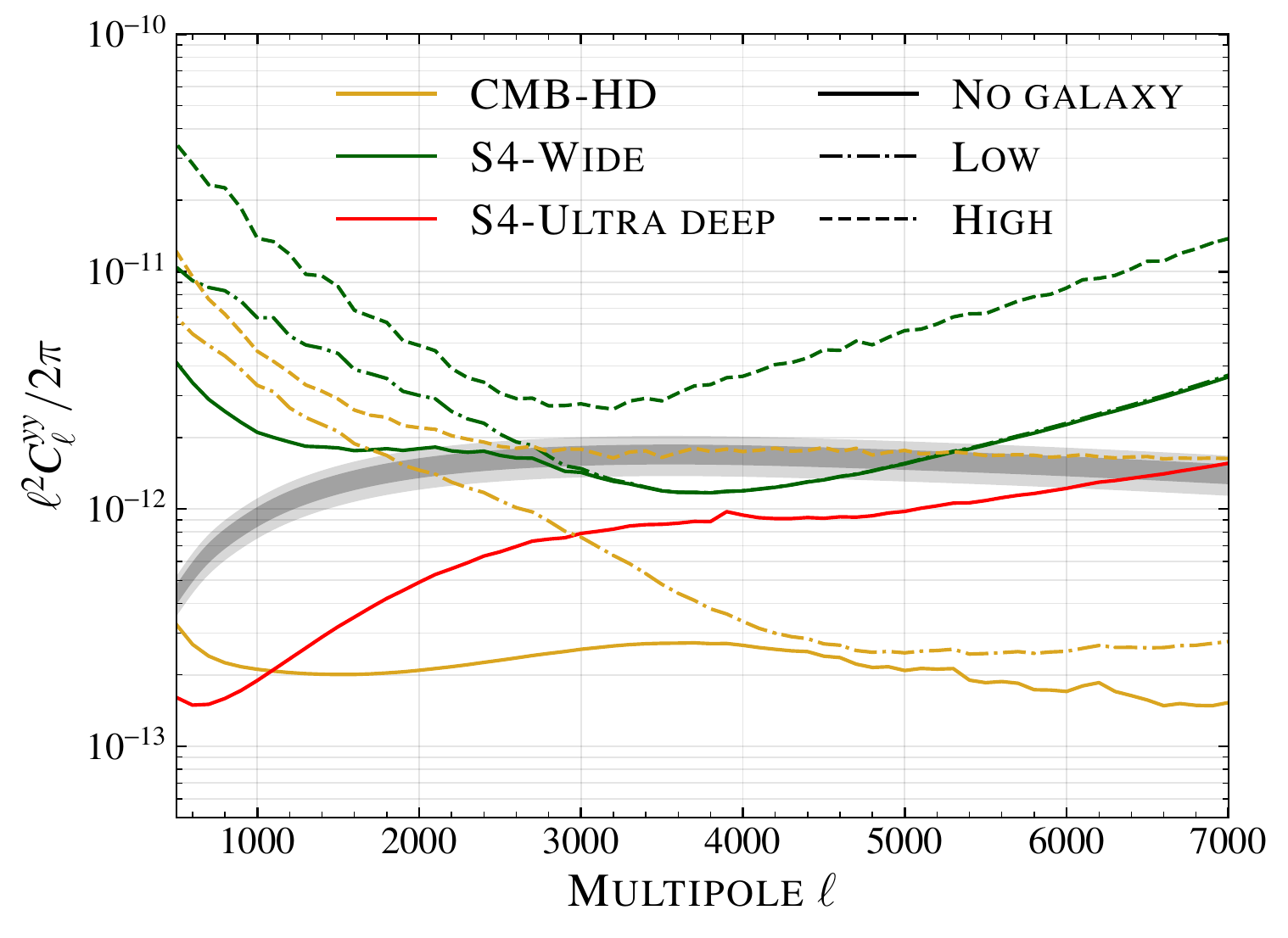}
\fi
\caption{
Residual ILC Compton-$y$ noise power spectra $\nlyy$ for the three surveys: \cmbhd{} in yellow, \sfour{} in green, and \sfourdeep{} in red. 
Solid curves correspond to noise curves without the inclusion of galactic signals. 
Dash-dotted and dashed lines are the residual noise curves in the clean (low) and dirty (high) galactic emission regions. 
The grey band shows the level of fiducial tSZ power spectrum along with $1\sigma, 2\sigma$ errors from \citet{george15}. 
Cluster detection sensitivity for CMB-S4 will be limited by residual CIB signals while the confusion noise from diffuse tSZ is the dominant source of variance for \cmbhd.
}
\label{fig_compton_nl_yy_ilc}
\end{figure}

Fig.~\ref{fig_compton_nl_yy_ilc} shows the residual power in the ILC Compton-$y$ maps ($\nlyy$) for \cmbhd{} (yellow), \sfour{} (green), and \sfourdeep{} (red). 
Solid lines in the figure correspond to noise estimates when galactic emission is not included. 
\cmbhd{} and \sfour{} experiments are expected to scan large sky areas ($\fsky = \fskywidefull$) and it is unrealistic to ignore galactic emission. 
Hence for \cmbhd{} (yellow) and \sfour{} (green) experiments we also show the noise curves in regions of low (dash-dotted) and high (dashed) galactic emission as discussed in \S\ref{sec_galactic_foregrounds}. 
Since \sfourdeep{} will observe a region with negligible galactic foregrounds \citep{cmbs4collab19}, we only show solid red line.

In the absence of galactic emission (solid curves), we find the noise level in \cmbhd{} maps to be much lower than both \sfour{} and \sfourdeep{} surveys. 
This is primarily due to reduced level of CIB signals expected in \cmbhd{} compared to CMB-S4. 
As described in \S\ref{sec_exgal_foregrounds}, note that the CIB power at 150 GHz for \cmbhd{} is lower than CMB-S4 by $17\times$ \citep{sehgal19}. 
The Compton-$y$ maps from CMB-S4 are fully dominated by residual CIB signals on small-scales. 
Residual CIB signals in Compton-$y$ maps can be lowered by nulling CIB signals assuming one or more spectral energy distributions with a constrained ILC \citep{madhavacheril20} or using partial ILC techniques \citep{bleem21}. 
This CIB reduction comes at the cost of higher noise depending on the choice of cleaning. 
We ignore this here but study the systematics in the recovered cluster tSZ signals due to emissions from DSFGs within clusters (\mbox{tSZ $\times$ CIB}) in \S\ref{sec_clus_corr_fg}. 

On large-scales, we note a change in noise trend and find noise in \cmbhd{} to be slightly higher than \sfourdeep. 
This is due to a higher atmospheric noise in \cmbhd{} as it will be located in Chile \citep{sehgal20} compared to \sfourdeep{} which will be observing from the South Pole. 
For \sfour{}, both atmospheric noise and residual CMB signals dominate the large-scale noise which is much higher than both \cmbhd{} and \sfourdeep{} surveys. 
Including information from \planck{} will improve the noise performance on large-scales but we ignore that as we are primarily interested in $\ell \gtrsim 3000$ for cluster detection. 

When galactic emission is included, as expected, the noise increases for both \cmbhd{} and \sfour{} surveys. 
For \sfour{}, adding low level of galactic emission (blue contour in Fig.\ref{fig_gal_emission}) only affects large-scale noise (green dash-dotted) as small-scales are dominated by residual CIB emission. 
When looking right through the galactic plane (red contour in Fig.\ref{fig_gal_emission}), residual noise (green dashed) is much higher on all scales. For \cmbhd{}, any level of galactic emission leads to an increased noise on all scales.

For reference, in grey band we show the fiducial tSZ power spectrum along with $1\sigma, 2\sigma$ errors from SPT measurement \citep{george15}. 
Comparing the grey band with noise curves, we can note that all the three surveys can map the peak of the tSZ power spectrum ($3000 \le \ell \le 4500$) with $\snr \ge 1$ \citep[accord][]{cmbs4collab19}. 

Besides the instrumental noise and foregrounds, another source of noise for cluster detection is the confusion noise arising due to the diffuse tSZ signal. 
Note that the noise curve $\nlyy$ is much lower than tSZ power spectrum for \cmbhd{}. 
While this indicates a high $\snr$ measurement of the tSZ power spectrum on all scales, it limits the sensitivity of cluster detection due to the tSZ confusion noise. 

\subsection{Baseline results}
\label{sec_baseline_results}
Our baseline results with no modifications to the cluster tSZ signal are presented in Figs.~\labelcref{fig_baseline_cluster_limiting_mass,fig_baseline_survey_completeness,fig_baseline_cluster_counts} and Tables~\ref{tab_cluster_counts} and \ref{tab_cluster_median_mass_redshift_lensing_constraints}. 

\subsubsection{Cluster detection sensitivity}
\label{sec_baseline_sensitivity}

\begin{figure}
\centering
\ifdefined\ApJsubmit
\includegraphics[width=0.48\textwidth, keepaspectratio]{limiting_masses_vs_z.pdf}
\else
\includegraphics[width=0.48\textwidth, keepaspectratio]{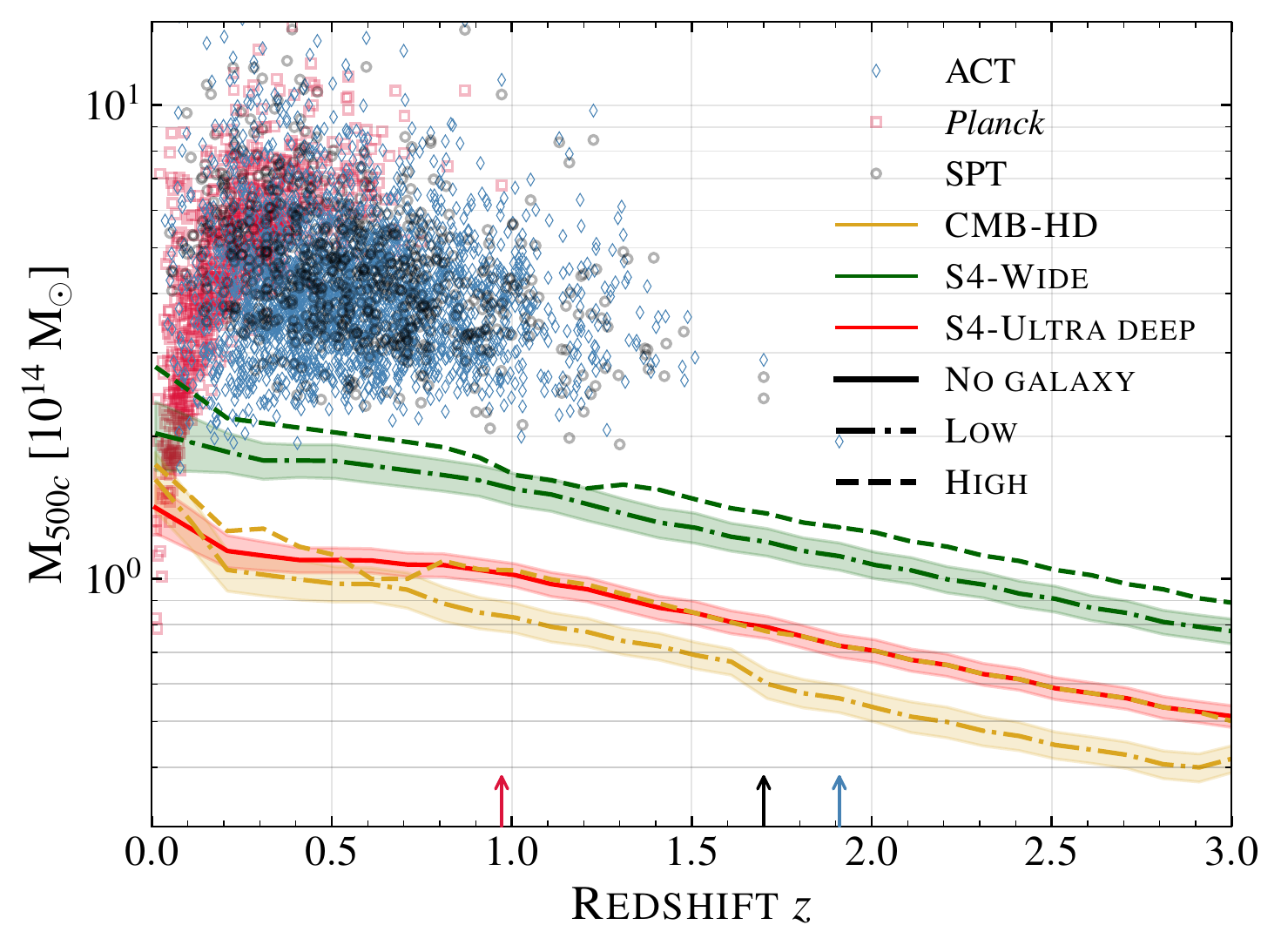}
\fi
\caption{
Cluster limiting mass threshold (\snr $\ge$ \snrlimit) as a function of redshift for future CMB surveys: \cmbhd{} in yellow, \sfour{} in green, and \sfourdeep{} in red. 
Different line styles correspond to the estimated sensitivity for different levels of galactic emission: solid for no, dash-dotted for low, and dashed for high galactic emissions respectively. 
See \S\ref{sec_galactic_foregrounds} and Fig.~\ref{fig_compton_nl_yy_ilc} for more details.
Masses and redshifts of clusters with \snr $\ge 4.5$ from currently available tSZ samples are also shown for comparison: 
\act{} \citep{hilton18, hilton21} as blue diamonds, \planck{} \citep{plancksz15} as red squares, and \spt{} \citep{bleem15, huang20, bleem20} as black circles. 
Arrows represent the redshift of the most distant cluster discovered in each survey. 
}
\label{fig_baseline_cluster_limiting_mass}
\end{figure}

Fig.~\ref{fig_baseline_cluster_limiting_mass} shows the redshift dependence of the minimum cluster mass required to satisfy the detection threshold criterion $\snr \ge \snrlimit$. 
For reference, we also mark the clusters detected at $\snr \ge 4.5$ from current surveys: ACT \citep{hilton18, hilton21} as blue diamond, \planck{} \citep{plancksz15} as red squares, and SPT \citep{bleem15, huang20, bleem20} as black circles. 
We present two curves for \cmbhd{} (yellow) and \sfour{} (green): dash-dotted and dashed curves correspond to sensitivity in regions of low and high levels of galactic emissions respectively. 
For \sfourdeep{}, since we do not inject any galactic emission, we only show the solid line containing no galactic foregrounds.

As expected based on the intuition from Fig.~\ref{fig_compton_nl_yy_ilc}, minimum detectable cluster mass is lowest for \cmbhd{} followed by \sfourdeep{} and \sfour{} surveys. 
The dominant source of variance for CMB-S4 surveys comes from the residual CIB contamination present in the ILC maps on small-scales. 
We tweaked CMB-S4's configuration to investigate if the residual CIB levels can be lowered further. 
To this end, we altered the noise level of bands in both CMB-S4 surveys by modifying the number of detectors in each band. 
We do not find any improvement which suggests that the current configuration listed in Table~\ref{tab_exp_specs} \citep{cmbs4collab19} is the most optimal for CMB-S4 cluster survey.

For \cmbhd, since the residual CIB level is expected to much lower than CMB-S4 in our setup \citep{sehgal19}, one could expect the cluster sensitivity to be much higher than CMB-S4. 
However, the confusion noise from diffuse tSZ \citep[accord][]{holder07} sets a noise floor hindering further improvement in sensitivity. 
Note that the grey signal band in Fig.~\ref{fig_compton_nl_yy_ilc} is much higher than the noise curves $\nlyy$ for \cmbhd{} in yellow. 
The tSZ confusion noise can be lowered by masking the detected clusters but we defer a detailed investigation of this to a future work. 

The sensitivity in regions of high galactic emission for \sfour{} is worse than the rest of the footprint by roughly 16\% at all redshifts. 
For \cmbhd{}, the degradation is $\sim28\%$ for clusters with $z \le 1$ and $\sim23\%$ overall. 
While $\lesssim 30\%$ \snr{} penalty is significant, we note that it is an optimistic estimate given that our model for the galactic emission power spectrum (\texttt{S0\_d0} dust and \texttt{S0\_s0} from pySM3 simulations) is a simple power-law. 
It ignores complexities like varying spectral or emissivity indices and non-Gaussianities which can introduce non-negligible biases to the cluster tSZ signal. 
As a result, we do not consider the clusters in regions of high galactic emission for subsequent analyses in this work. 

While not shown in Fig.~\ref{fig_baseline_cluster_limiting_mass}, in the absence of galactic emission, cluster limiting masses reduce by $\sim7\%$ compared to regions with low levels of galactic emission in the \sfour{} footprint. 
A significant fraction of this $\snr$ degradation in the presence of galactic emission is for nearby clusters in agreement with excess large-scale noise, green solid vs green dash-dotted curves, in Fig.~\ref{fig_compton_nl_yy_ilc}. 
For all the three surveys, the spike at low redshift in Fig.~\ref{fig_baseline_cluster_limiting_mass} is because we limit \snr{} calculation to \mbox{$\theta_\mathrm{max}=\thetamax$}. 
While $\thetamax >>\thetavir $ for  clusters with $z \gtrsim 0.5$ and hence optimal, our choice of $\theta_\mathrm{max}$ does not fully encompass the cluster signal for nearby clusters and hence reduces their \snr. 

The reason for the decrease in the minimum detectable mass with redshift is two-folded. 
At low redshifts, the cluster SNR degrades because of residual contamination from atmospheric noise and CMB. 
At high redshifts, according to self-similar evolution of clusters \citep{kaiser86}, a cluster with a given mass will have a higher temperature and hence a higher tSZ signal compared to its low redshift counterpart. 
This leads to an increase in cluster SNR when going from low to high redshifts.

\subsubsection{Survey completeness}
\label{sec_completeness}

\begin{figure}
\centering
\ifdefined\ApJsubmit
\includegraphics[width=0.48\textwidth, keepaspectratio]{survey_completeness_estimates_Ysz.pdf}
\else
\includegraphics[width=0.48\textwidth, keepaspectratio]{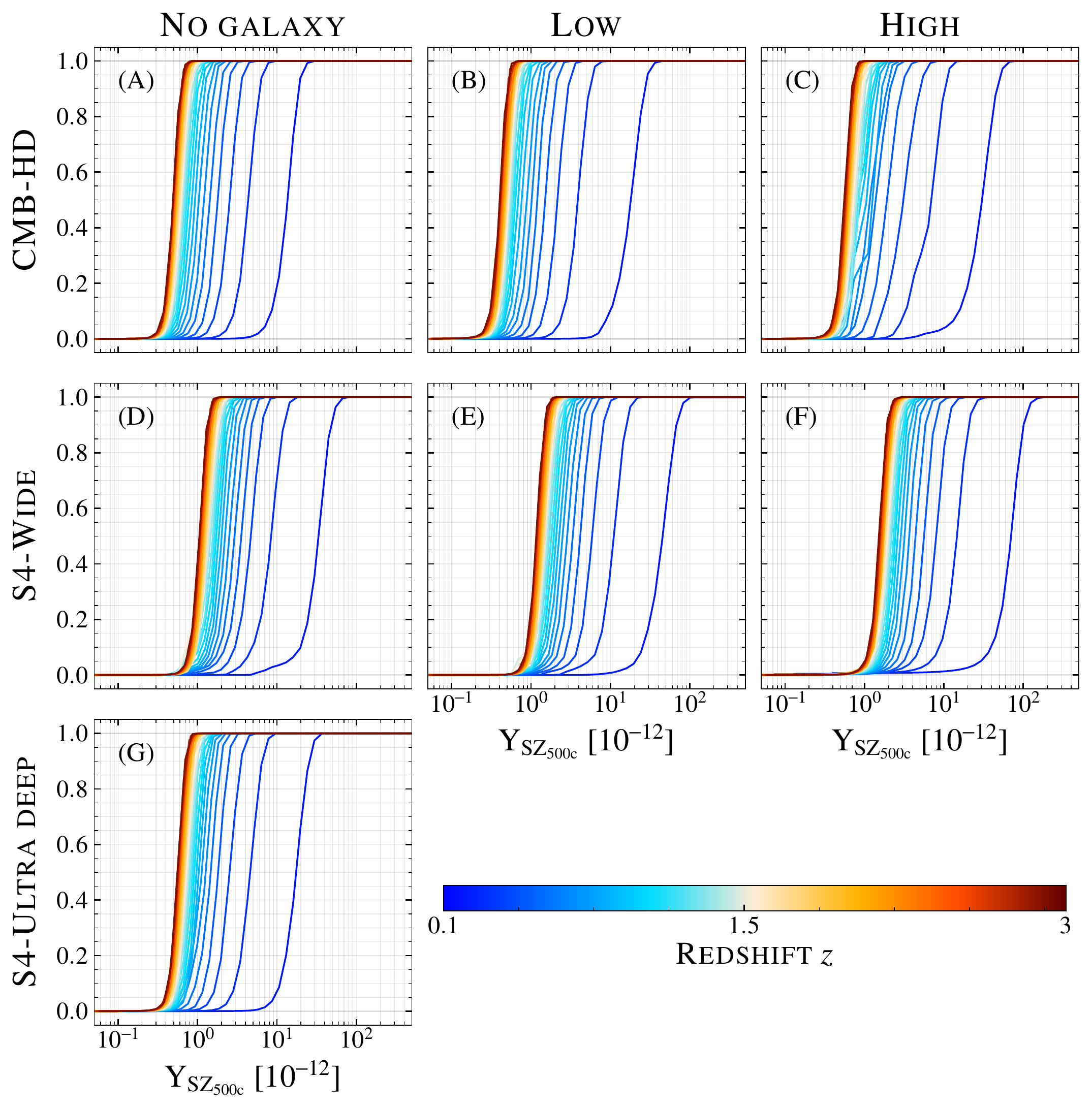}
\fi
\caption{
Cluster survey completeness as a function of integrated $\Yvir$ signal for the three surveys with different levels of galactic emission. 
As expected, the curves indicate that surveys with lower instrumental noise and galactic foregrounds will allow us to detect clusters with lower $\Yvir$ signals. 
The redshift trend is due to self-similar evolution of clusters and the residual contamination from CMB and atmospheric noise in the Compton-$y$ maps.
}
\label{fig_baseline_survey_completeness}
\end{figure}

Sensitivity can also be expressed in terms of cluster survey completeness as \citep{plancksz15, alonso16} 
\begin{equation}
\chi(\Yvir) = \frac{1}{2} \left[ 1 + {\rm erf} \left( \frac{\Yvir^{\rm true} - q_{\rm lim} \sigma_{\Yvir}} {\sqrt{2}\sigma_{\Yvir}}\right)\right],
\end{equation}
where $q_{\rm lim} = \snrlimit$ is the detection threshold, $\sigma_{\Yvir}$ is the measurement uncertainty of the integrated $\Yvir$ signal estimated in \S\ref{sec_ml_fitting}, and $\Yvir^{\rm true}$ is the true $\Yvir$ flux. 
Cluster limiting mass as a function of redshift shown in Fig.~\ref{fig_baseline_cluster_limiting_mass} represents 50\% survey completeness.
In Fig.~\ref{fig_baseline_survey_completeness}, we show the completeness as a function of $\Yvir$ for all the three surveys for different levels of galactic emission. 
Here, $\Yvir$ is the integrated Compton signal within the virial radius $R_{500}$.
Colors represent cluster redshift with $z = 0$ in blue and $z = 3$ in red. 
Higher experimental sensitivity will result in steeper curves. 
It will also push the curves to the left enabling detection of clusters with a lower $\Yvir$ signal.
This is evident from the figure where we note that curves move from lower to higher values of $\Yvir$ for increasing levels of galactic emission (left to right). 
The slope of individual lines also decrease in the same order. 
We note the same pattern when going from low-noise to high-noise surveys (\cmbhd{} $\rightarrow$ \sfourdeep{} $\rightarrow$ \sfour) and also from low to high cluster redshifts (blue to red). 
The redshift trend is because of: (a) SNR degradation for low redshift clusters due to residual contamination from atmospheric noise and CMB and (b) SNR improvement due to self-similar evolution for higher redshift clusters. 
The significant \snr{} penalty for lowest redshifts $z \lesssim 0.3$ is due to the hard cutoff \mbox{$\theta_\mathrm{max}=\thetamax$} used for \snr{} calculation. 
See \S\ref{sec_baseline_sensitivity} for more discussion.

Based on these results, we find that \sfour{} shall detect (at $5\sigma$) all galaxy clusters with an integrated Compton $\Yvir \ge 10^{-12}$ at $z \ge 1.5$ over the large area survey footprint ($\fsky = \fskywideclean$) as shown in panel (E) of Fig.\ref{fig_baseline_survey_completeness}.  
Furthermore, \sfourdeep{} shall detect (at $5\sigma$) all galaxy clusters with an integrated Compton $\Yvir \ge 5\times10^{-13}$ at $z \ge 1.5$ over the de-lensing survey footprint ($\fsky = \fskydeepfull$) shown in the bottom panel (G). 
The sensitivity of \cmbhd{} is roughly similar to \sfourdeep{} but over a large region $\fsky = \fskywideclean$ of sky as shown in panel (B).

\subsubsection{Cluster counts}
\label{sec_baseline_cluster_counts}

\begin{deluxetable*}{| l | c | c | c | c | c | c | c | c | c |}
\def\arraystretch{1.2}
\tablecaption{Forecasted number of clusters from future SZ surveys with $\snr \ge \snrlimit$ in regions with different levels of galactic emission.}
\label{tab_cluster_counts}
\tablehead{
\multirow{2}{*}{Experiment} & \multicolumn{3}{c|}{Baseline footprint} & \multicolumn{3}{c|}{Dirty footprint} & \multicolumn{3}{c|}{Full footprint}\\
\cline{2-10}
& $\fsky$ & Total & $z \ge 2$& $\fsky$ & Total & $z \ge 2$& $\fsky$ & Total & $z \ge 2$
}
\startdata
\cmbhd{} & \multirow{2}{*}{\fskywideclean} & \howmanyclustersforcmbhdbaseline & \howmanyhighzclustersforcmbhdbaseline & \multirow{2}{*}{\fskywidedirty} & \howmanyclustersforcmbhddirty & \howmanyhighzclustersforcmbhddirty & \multirow{2}{*}{\fskywidefull} & $\howmanyclustersforcmbhdfull$ & $\howmanyhighzclustersforcmbhdfull$ \\
\cline{1-1}\cline{3-4}\cline{6-7}\cline{9-10}
\sfour{} & & \howmanyclustersforsfourbaseline & \howmanyhighzclustersforsfourbaseline & & \howmanyclustersforsfourdirty & \howmanyhighzclustersforsfourdirty & & $\howmanyclustersforsfourfull$ & $\howmanyhighzclustersforsfourfull$ \\\hline\hline
\sfourdeep{} & \fskydeepfull & \howmanyclustersforsfourdeepbaseline & \howmanyhighzclustersforsfourdeepbaseline & \multicolumn{3}{c|}{-} & \fskydeepfull & \howmanyclustersforsfourdeepbaseline & \howmanyhighzclustersforsfourdeepbaseline \\\hline\hline
\enddata
\end{deluxetable*}

\begin{deluxetable}{| l | c | c | c | c | c |}
\tablecaption{Median mass and redshift of cluster sample from the future SZ surveys in their baseline footprint along with the CMB-cluster lensing mass constraints.}
\label{tab_cluster_median_mass_redshift_lensing_constraints}
\def\arraystretch{1.2}
\tablewidth{.8\columnwidth}
\tablehead{
\multirow{2}{*}{Experiment} & \multirow{2}{*}{$z^{\rm med}$} & \multicolumn{2}{c|}{$\mvir^{\rm med}\ [\munitssimple]$} & \multicolumn{2}{c|}{$\Delta M^{\rm stack}_{\rm lens}$ [$\munitssimple$]}\\
\cline{3-6}
& & All & $z \ge 2$ & All & $z \ge 2$
}
\startdata
\cmbhd{} & \medianzcmbhd & \medianmasscmbhd & \medianmasshighzcmbhd & \lensingmasserrorcmbhd & \lensingmasserrorhighzcmbhd \\\hline
\sfour{} & \medianzsfour & \medianmasssfour & \medianmasshighzsfour & \lensingmasserrorsfour & \lensingmasserrorhighzsfour \\\hline\hline
\sfourdeep{} & \medianzsfourdeep & \medianmasssfourdeep & \medianmasshighzsfourdeep & \lensingmasserrorsfourdeep & \lensingmasserrorhighzsfourdeep \\\hline\hline
\enddata
\end{deluxetable}

\begin{figure}
\centering
\ifdefined\ApJsubmit
\includegraphics[width=0.45\textwidth, keepaspectratio]{clus_counts.pdf}
\else
\includegraphics[width=0.45\textwidth, keepaspectratio]{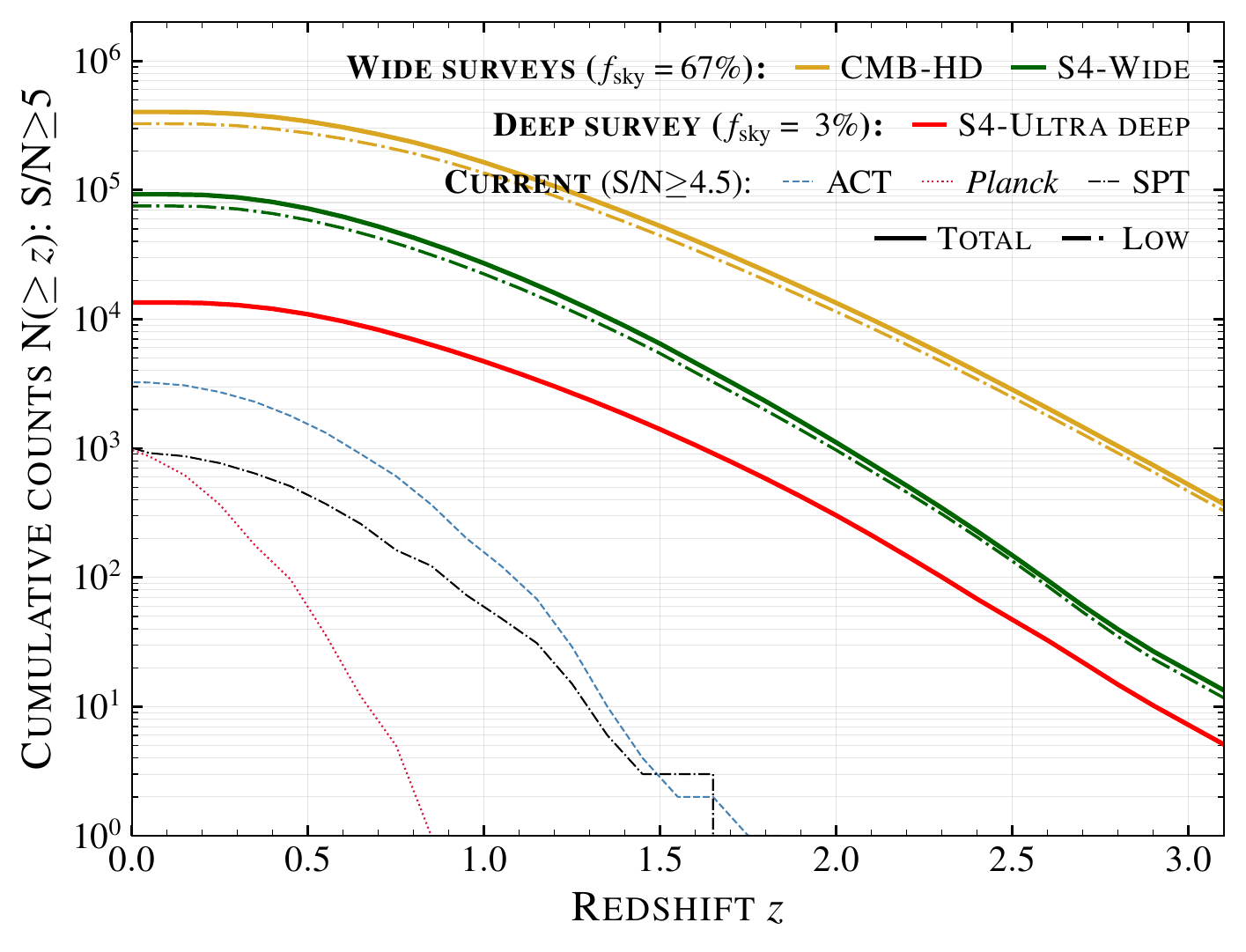}
\fi
\caption{
Cumulative cluster redshift distribution for future (current) SZ samples are shown as thick (thin) curves. 
For \cmbhd{} (yellow) and \sfour{} (green), we show two curves: dash-dotted corresponds to clusters expected in the baseline (low galactic emission) footprint ($\fsky = \fskywideclean$) and solid corresponds to the total clusters expected in the combined low and high galactic emission regions. 
CMB-S4 is expected to detect close to 1000 (350) clusters at $z \ge 2$ in the baseline footprint $\fsky = \fskywideclean\ (\fskydeepfull)$ with the \sfour{} (\sfourdeep) survey. 
The number of $z\ge 2$ is more than an order of magnitude larger for \cmbhd{} compared to CMB-S4. 
The enormous improvement in sensitivity of high redshift clusters for future surveys compared to current experiments (thin lines) is evident from the curves.
}
\label{fig_baseline_cluster_counts}
\end{figure}

We present cumulative redshift distribution of clusters expected from the three surveys in Fig.~\ref{fig_baseline_cluster_counts}: \cmbhd{} in yellow, \sfour{} in green, and \sfourdeep{} in red. 
Cluster counts are obtained by sampling \citet{tinker08} HMF using the MC sampling approach discussed in \S\ref{sec_monte_carlo_sampling}.
Dash-dotted lines for \cmbhd{} and \sfour{} correspond to clusters expected from regions with low galactic foregrounds. 
Solid curves are the total number of clusters from the full footprint (i.e:) combination of both low and high galactic emission regions and the split between the two regions can be found in Table~\ref{tab_cluster_counts}.
Like in previous sections, galactic foregrounds are absent for \sfourdeep. 
For comparison, we show the currently available SZ cluster samples (\snr $\ge 4.5$) from \act{} \citep{hilton18, hilton21} as blue dashed, \planck{} \citep{plancksz15} as red dotted, and \spt{} \citep{bleem15, huang20, bleem20} as black dash-dotted curves. 

\sfour{} shall detect close to 75,000 clusters in the baseline footprint ($\fsky = \fskywideclean$) while the \sfourdeep{} will obtain $\sim14,000$ clusters in the CMB-S4's delensing footprint ($\fsky = \fskydeepfull$). 
While most of the low redshift $z \lesssim 1$ clusters will be part of the LSST or eROSITA cluster samples \citep{lsst09, erosita12}, the redshift independent property of the tSZ signal will open the unique high redshift discovery space for future CMB surveys. 
For example, \sfour{} (\sfourdeep) is expected to detect 1000 (350) clusters at $z \ge 2$. 
The number of clusters expected from \cmbhd{} is $\times3$ more than \sfour. In the high redshift regime $z \ge 2$, the expected number for \cmbhd{} is more than an order of magnitude higher than CMB-S4. 
In Table~\ref{tab_cluster_median_mass_redshift_lensing_constraints}, we give the median masses and redshifts of clusters in the baseline footprint from all the three surveys. 
Average lensing mass estimates of the cluster sample using both temperature and polarization CMB-cluster lensing is also given in the table. 
While CMB temperature returns a higher lensing \snr{} for \sfour, we find polarization channel to dominate the \snr{} for \sfourdeep{}. 
This is due to higher noise floor set by foreground signals along with additional strategies used to mitigate foreground-induced bias in temperature based lensing reconstruction. 
The same is true for \cmbhd{} but to a much lower extent since the variance from CIB is highly suppressed for \cmbhd{} \citep{sehgal19}. 
We report median mass and lensing estimates for both the full sample and also for clusters with $z \ge 2$. 

\subsection{Change in sensitivity due to changes in virialization}
\label{sec_sensitivity_vir_model_dependence}

\begin{figure}
\centering
\ifdefined\ApJsubmit
\includegraphics[width=0.45\textwidth, keepaspectratio]{limiting_masses_vs_z_for_various_alpha.pdf}
\else
\includegraphics[width=0.45\textwidth, keepaspectratio]{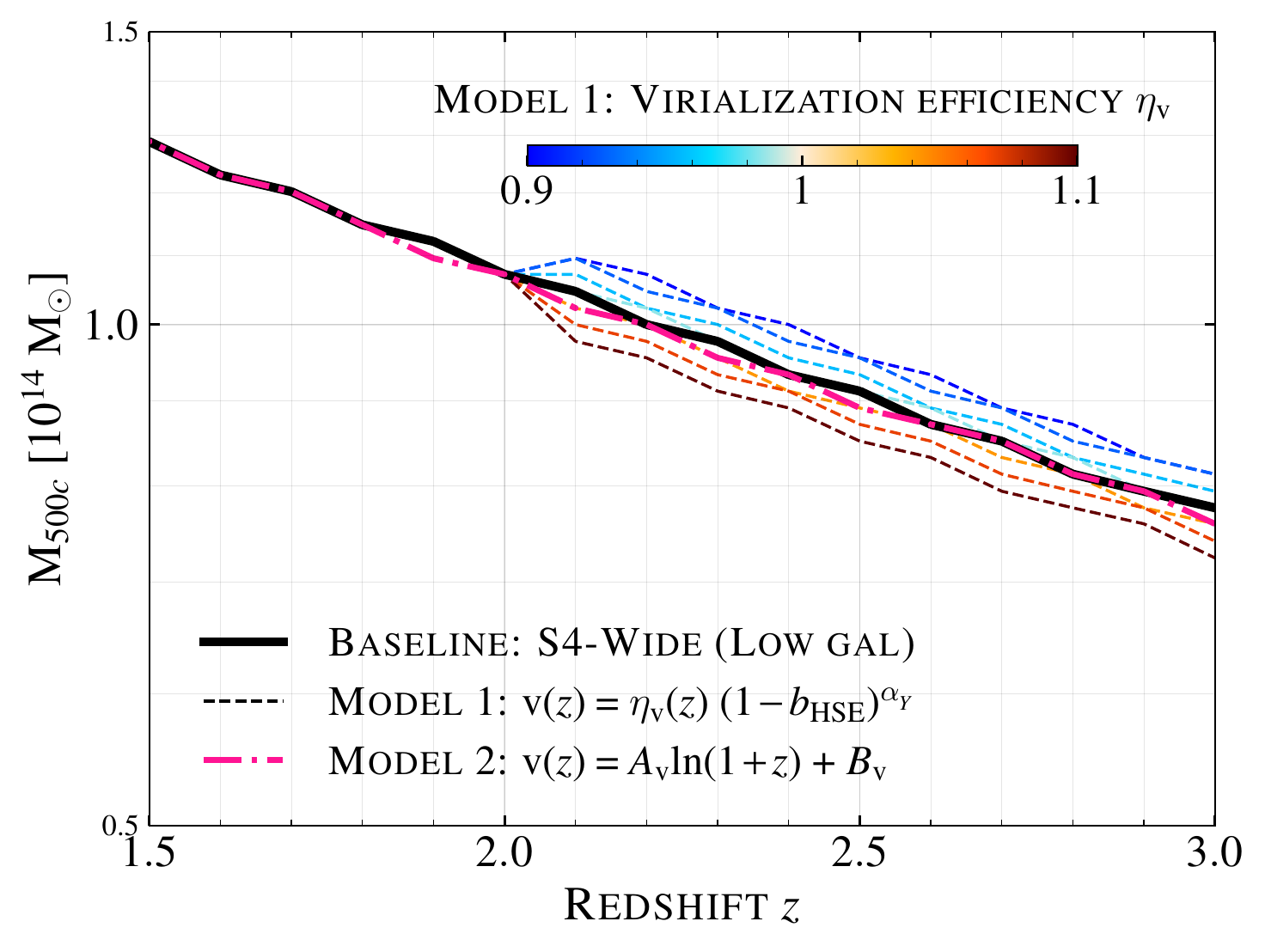}
\fi
\caption{Dependence of cluster detection sensitivity on the virialization model. 
Black curve is the baseline case for \sfour{} survey in the region with low galactic emission, same as green dash-dotted curve in Fig.~\ref{fig_baseline_cluster_limiting_mass}.
Thin dashed curves for model 1 with virialization efficiency parameter ranging from  $\vireffeta \in$ [0.9 (blue), 1.1 (red)]. 
Thick pink dash-dotted curve is for model 2 with $\virslope = \virslopeval$ and $\virintercept = \virinterceptval$.
}
\label{fig_s4_wide_limiting_mass_redshift_diff_vir_eff_models}
\end{figure}

Modifying cluster virialization alters cluster tSZ signal from clusters which in turn affects the tSZ \snr{}. 
This is illustrated using the change in the minimum detectable cluster mass as a function of redshift in Fig.~\ref{fig_s4_wide_limiting_mass_redshift_diff_vir_eff_models} for \sfour{} with low levels of galactic emission. 
Thick solid black line is the baseline curve, same as green dash-dotted curve in Fig.~\ref{fig_baseline_cluster_limiting_mass}. 
Thin dashed curves are for model 1 when we vary the virialization efficiency from $\vireffeta \in [\vireffetasmall, \vireffetalarge]$ based on Eq.(\ref{eq_vir_model_1}) and Eq.(\ref{eq_eta_z}). 
As expected, the minimum detectable masses decreases for $\vireffeta > \vireffeta^{\rm fid} (= \vireffetafid)$.
The number of high redshift $z \ge 2$ clusters from \sfour{} drop (increase) by $\times2$ for $\vireffeta = \vireffetasmall\ (\vireffetalarge)$ compared to $\howmanyhighzclustersforsfourbaseline$ clusters for the fiducial value $\vireffeta^{\rm fid} = \vireffetafid$ (see Table~\ref{tab_cluster_counts}). 

Thick pink dash-dotted curve is for model 2 based on Eq.(\ref{eq_vir_model_2}). 
It is similar to our baseline case (black) and we get roughly a 10\% overall increase in the number of clusters consistent with the trend in Fig.~\ref{fig_s4_wide_limiting_mass_redshift_diff_vir_eff_models}. 

\subsection{Fisher forecasts}
\label{sec_forecasts}

\begin{figure*}
\centering
\ifdefined\ApJsubmit
\includegraphics[width=.9\textwidth, keepaspectratio]{triangle_alpha-intercept_vir-m_nu-one_minus_hse_bias-slope_vir-w0_100mciters.pdf}
\else
\includegraphics[width=.9\textwidth, keepaspectratio]{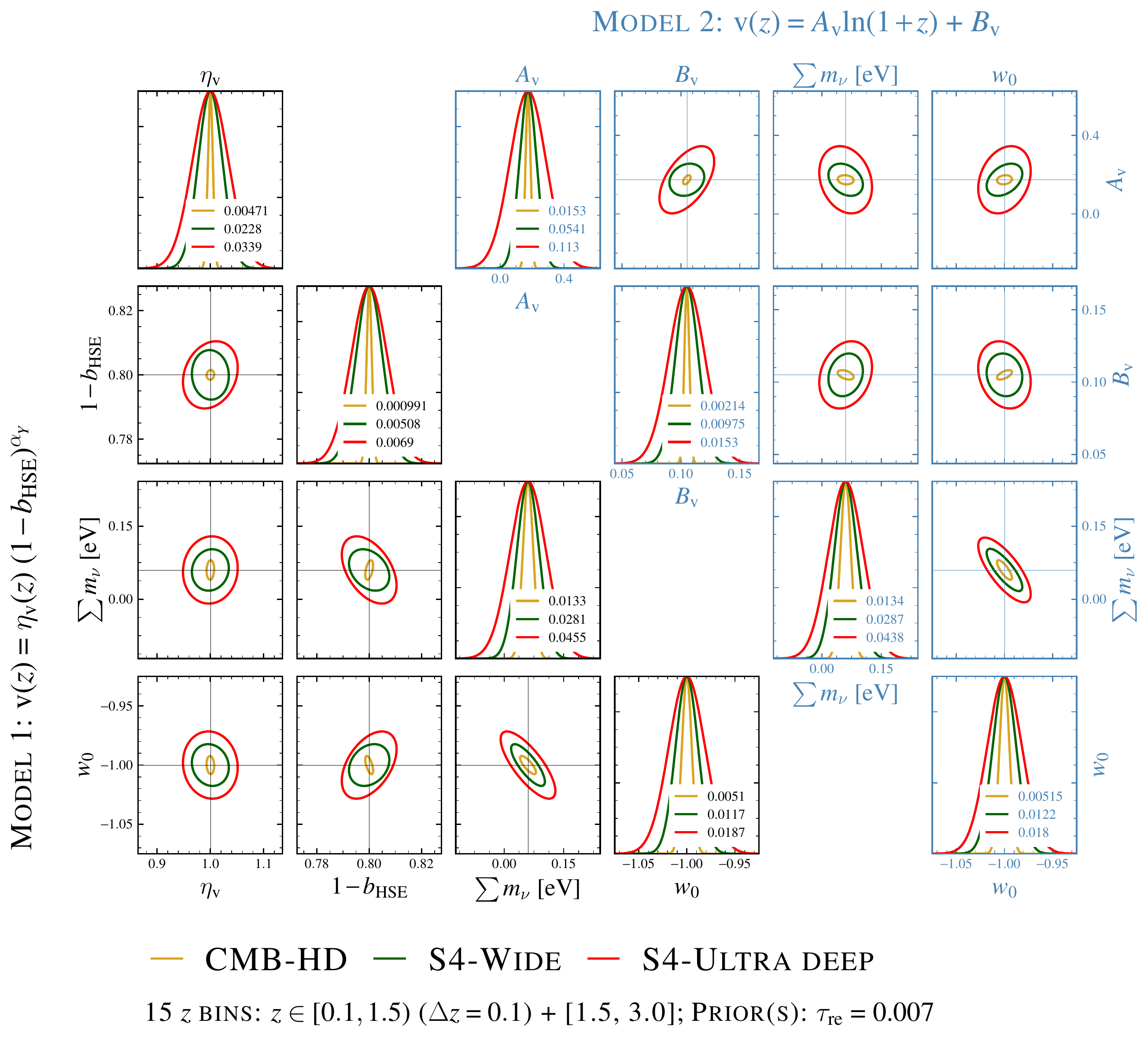}
\fi
\caption{
Marginalized Fisher constraints (68\% CL) obtained by combining information from primary CMB spectra (TT/EE/TE) and cluster counts $\binnedcluscounts$: \cmbhd{} is in yellow, \sfour{} in green, and \sfourdeep{} in red. 
We use a \planck-like prior $\sigma(\taure) = 0.007$ for all surveys. 
Both surveys can reduce the uncertainty on the dark energy equation of state parameter $\sigma(\wde)$ to $\lesssim 1\%$. 
Combining primary CMB with clusters will also enable $\sim 2.5-4.5\sigma$ detection of the neutrino masses. 
Lower and upper diagonal represent cluster virialization models 1 and 2 respectively. 
{\it \mbox{Model 1:}} Cluster virialization efficiency $\vireffeta$ can be constrained to an accuracy of $2-4\%$ level by \sfour{} and \sfourdeep{} while \cmbhd{} can provide sub-percent level constraints. 
All surveys provide $<1\%$ constraints on the HSE bias parameter.
{\it \mbox{Model 2:}} CMB-S4 can provide 33\% and $\sim$ 4\% constraints on $\virslope$ and $\virintercept$ parameters while \cmbhd{} reduces the uncertainties on both parameters by $\ge \times 3$. 
Errors on other parameters do not change significantly between the two cluster virialization models. 
}
\label{fig_constraints_s4_cmbhd}
\end{figure*}

\begin{figure*}
\centering
\ifdefined\ApJsubmit
\includegraphics[width=0.9\textwidth, keepaspectratio]{triangle_cmb_cluster_joint_alpha-intercept_vir-m_nu-one_minus_hse_bias-slope_vir-w0_100mciters.pdf}
\else
\includegraphics[width=0.9\textwidth, keepaspectratio]{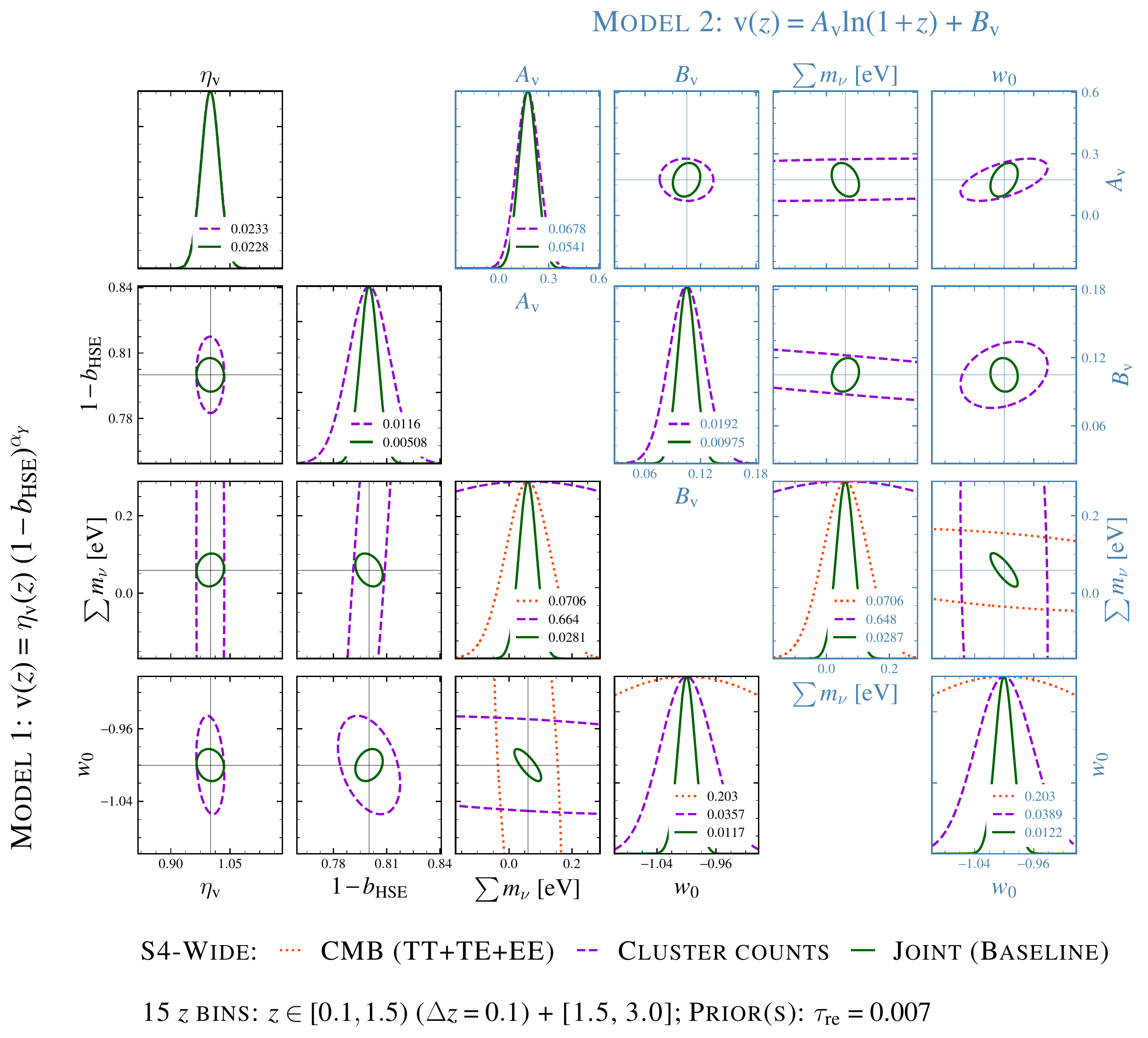}
\fi
\caption{
Individual constraints (68\% CL) from CMB TT/EE/TE spectra (orange dotted) and cluster counts (purple dashed) are shown. 
The combination of the two, our baseline setup, are shown as green solid curves. 
The nearly-orthogonal degeneracy directions of CMB spectra and cluster counts on structure growth parameters provide excellent joint constraints compared to either of them individually on $\sigma(\summnu)$ and $\sigma(\wde)$. 
Only \sfour{} is shown. 
}
\label{fig_constraints_s4wide_cmbclusterjoint}
\end{figure*}

\begin{figure*}
\centering
\ifdefined\ApJsubmit
\includegraphics[width=0.9\textwidth, keepaspectratio]{triangle_no_cmblensing_alpha-intercept_vir-m_nu-one_minus_hse_bias-slope_virw0_100mciters.pdf}
\else
\includegraphics[width=0.9\textwidth, keepaspectratio]{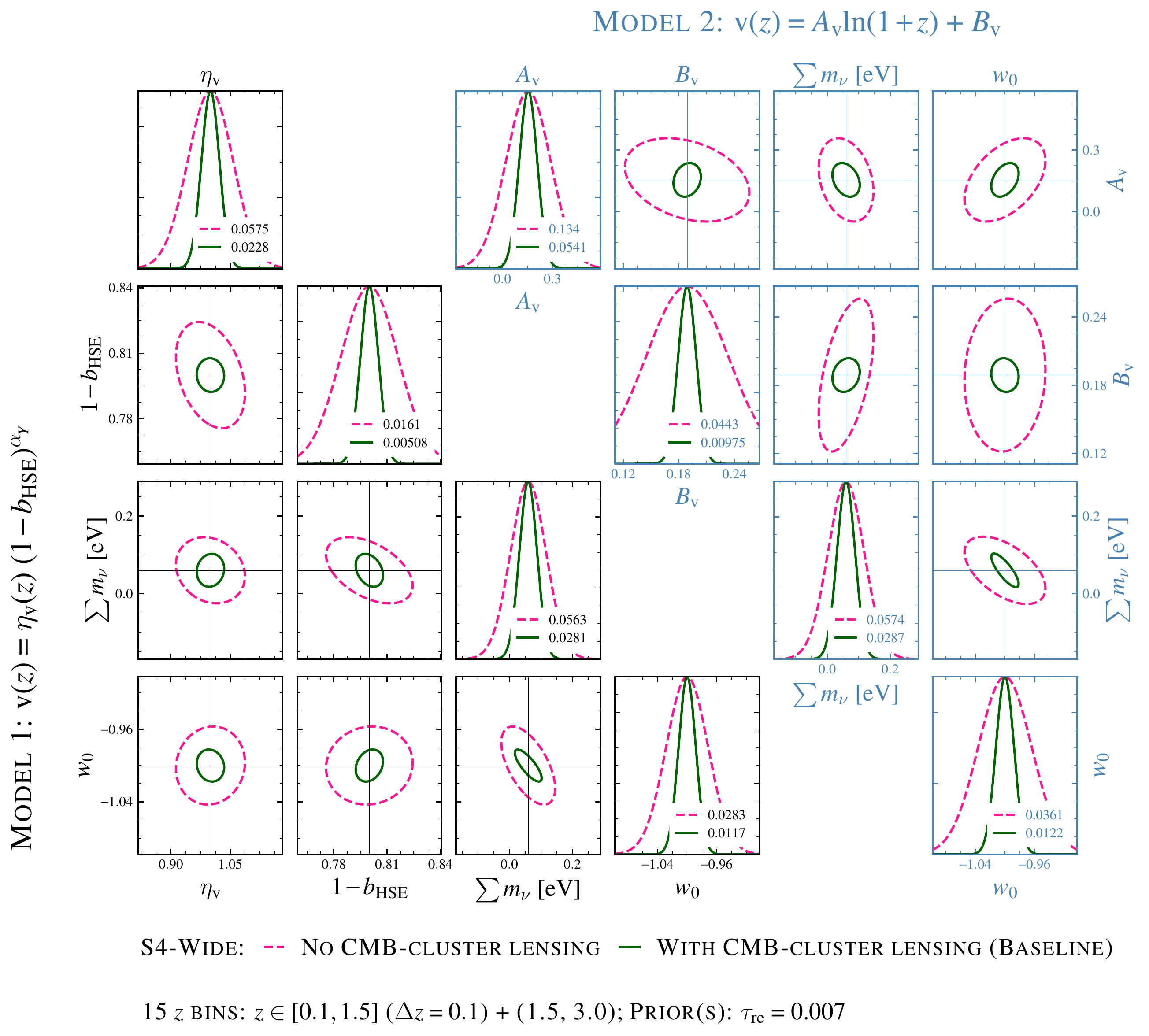}
\fi
\caption{Importance of CMB-cluster lensing mass calibration for cluster counts. Green curves are the same as left panel while the pink dashed curves are constraints obtained without CMB-cluster lensing information. Ignoring lensing mass calibration degrades the constraints significantly for all parameters. 
Ellipses are 68\% CL regions and only \sfour{} is shown.
}
\label{fig_constraints_s4wide_masscalib}
\end{figure*}

Now we turn to parameter constraints using the Fisher formalism presented in \S\ref{sec_fisher_formalism}. We combine $\binnedcluscounts$ with primary CMB information from the three surveys to forecast standard errors on the parameters governing one of the two virialization models along with \mbox{\mbox{$\YszM$}} scaling relation (Eq.~\ref{eq_Ysz_mass}) and cosmological parameters. 
For cosmology, we focus on two-parameter extension to $\lcdm$ to include sum of neutrino masses $\summnu$ and dark energy equation of state $\wde$. 
Unless otherwise stated, {\it cluster counts} $\binnedcluscounts$ in the rest of this section includes temperature and polarization based CMB-cluster lensing mass calibration. 
The baseline redshift binning adopted was $\Delta z = 0.1$ for $0.1 \le z < 1.5$ and one massive redshift bin for all high redshift clusters $1.5 \le z \le 3 $. 
\planck-like prior has been assumed for optical depth to reionization $\sigma(\taure) = 0.007$.

We also look into the following: (a) individual constraints from primary CMB and cluster counts, (b) importance of CMB-cluster lensing based mass calibration, (c) impact of high redshift clusters and redshift binning and (d) the effect of $\taure$ prior. 
These checks are limited to \sfour{} only. 

\subsubsection{Cosmology and cluster virialization model}
\label{sec_vir_model_cosmology_constraints}

In Fig.~\ref{fig_constraints_s4_cmbhd}, we present the marginalized constraints (68\% CL) on $\theta \in [\vireffeta,\ \hsebias,\ \virslope,\ \virintercept,\ \summnu,\ \wde]$ from all the three surveys: \cmbhd{} in yellow, \sfour{} in green, and \sfourdeep{} in red. 
Lower and upper diagonal correspond to constraints for cluster virialization models 1 and 2 respectively.

The CMB-S4 and \cmbhd{} experiments can provide stringent constraints on dark energy equation of state $\sigma(\wde)$ and the sum of neutrino masses. 
We obtain 1-2\% on $\sigma(\wde)$ from CMB-S4: 1.2\% (1.8\%) from \sfour{} (\sfourdeep) and 1\% jointly from both CMB-S4 surveys. 
\cmbhd{} will provide sub-percent (0.5\%) level constraints on $\wde$.
For neutrino masses, we obtain $\sigma(\summnu) = 28\ \mev\ (45\ \mev)$ from \sfour{} (\sfourdeep), $23\ \mev$ jointly from both, and $13\ \mev$ from \cmbhd{} enabling a $\sim 2.5-4.5\sigma$ detection of the sum of neutrino masses from both CMB-S4 and \cmbhd{} assuming a normal hierarchy lower limit of $60\ \mev$. 
Although not shown in the figures, both experiments provide $\lesssim 1\%$ constraints on the scalar fluctuation amplitude $A_{s}$, Hubble parameter $\sigma(h)$, and dark matter density $\sigma(\omchsq)$. 
Adding large-scale information from \planck{} has negligible impact on the constraints from \sfour{} and \cmbhd{}, while it improves the cosmological constraints from \sfourdeep{} by $5-10\%$.

Results for the first virialization model in Eq.(\ref{eq_vir_model_1}) are shown in the lower diagonal of Fig.~\ref{fig_constraints_s4_cmbhd}. 
We find that CMB-S4 clusters can help constrain $\sigma(\vireffeta)$ at $2-4\%$ level while \cmbhd{} can provide sub-percent level constraints. 
However, note that this assumes we have 100\% knowledge about the gastrophysics of low redshift clusters which is not fully true but has been rapidly advancing \citep[see recent review by][]{mroczkowski19}.
Both experiments can provide sub-percent constraints on the HSE bias $1 - \hsebias$. 
While $\vireffeta$ only modifies tSZ signal of clusters at $z \ge 2$ and is only constrained by them, low redshift clusters are also important in breaking the degeneracies between other cosmological/scaling relation parameters and $\vireffeta$. 
For example, if we only consider clusters at $z \ge 2$, $\vireffeta$ is highly degenerate parameters like $\hsebias$ or $\sigma_{8}$ and adding low redshift information almost entirely breaks the degeneracies with other parameters. 

The upper diagonal of Fig.~\ref{fig_constraints_s4_cmbhd} shows the results for the second cluster virialization model in Eq.(\ref{eq_vir_model_2}). 
In this case, we obtain $\sigma(\virslope) = 0.05\ (0.1)$ from \sfour{} (\sfourdeep) for the redshift evolution parameter of the virialization, corresponding to $\sim33\%$ jointly from the two CMB-S4 surveys. 
For $\sigma(\virintercept)$, we get 5\% and 8\% from the two CMB-S4 surveys and $\sim4\%$ jointly. 
\cmbhd{} will reduce the measurement uncertainty on these two parameters by more than a $\times3$. 

Modifying cluster virialization from model 1 to model 2 does not introduce statistically significant differences in other parameter constraints as can be seen by comparing the lower and upper diagonals in Fig.~\ref{fig_constraints_s4_cmbhd}.

\subsubsection{Observable-mass scaling relation}
\label{sec_observable_mass_scaling_relation_constraints}

The scatter in the \mbox{$\YszM$} scaling relation is constrained to $13-16\%$ level by CMB-S4 and 2\% by \cmbhd. 
The $1\sigma$ errors on mass and redshift evolution parameters of the relation are $\sigma(\alphay) = 0.01$, $\sigma(\betay) = 0.02$, and $\sigma(\gammay) = 0.02$ for \sfour. 
When switching from model 1 to model 2, we note a strong degeneracy between $\virslope$ and $\gammay$ since they both probe the redshift evolution of the tSZ signal. 
The mass and redshift evolution parameters of the log-normal scatter ($\alphasigmalogy$ and $\gammasigmalogy$) are an order of magnitude worse. 
The numbers are similar or sometimes slightly better for \sfourdeep{} which is because of a better lensing \snr{} per cluster for \sfourdeep. 
We note that \cmbhd{} can improve constraints on the \mbox{$\YszM$} scaling relation parameters by roughly an order of magnitude compared to CMB-S4.

\subsubsection{CMB vs cluster counts}
\label{sec_cmb_cluster_constraints}
In Fig.~\ref{fig_constraints_s4wide_cmbclusterjoint}, we present the constraints (68\% CL) separately from CMB TT/EE/TE (orange dotted) and cluster counts (purple dashed) for \sfour. 
Green solid curves correspond to the joint CMB and cluster counts constraints. 
As before, lower and upper diagonals correspond to virialization models 1 and 2. 
CMB spectra is insensitive to cluster virialization parameters ($\vireffeta, \hsebias, \virslope, \virintercept$) and hence orange dotted curves are not shown for those parameters. 
However, CMB still helps in constraining them by breaking degenaracies with other parameters. 
That is the reason for the difference between purple dashed (cluster counts) and green solid (joint constraints) curves. 
For other parameters, CMB spectra adds minimal to modest level of information compared to clusters.
However, since CMB and cluster counts prefer different nearly-orthogonal degeneracy directions, the joint constraints offer remarkable improvements compared to either of them individually.
For example, constraints on the sum of neutrino masses 
improves by $\times2.5$, from $\sigma(\summnu) = 70\ \mev$ to $28\ \mev$, when adding cluster counts to primary CMB spectra.

\subsubsection{Importance of lensing masses}
\label{sec_lensing_no_lensing_constraints}
CMB-cluster lensing based mass calibration is critical to obtain the results described above. 
To highlight the importance, we present constraints with (green solid) and without (pink dashed) CMB-cluster lensing information in Fig.~\ref{fig_constraints_s4wide_masscalib}. 
When CMB-cluster lensing in excluded, we simply bin clusters in tSZ \snr{} $q$ and redshift $z$: $\binnedcluscountsnolensing$. 
Since both cosmological and \mbox{\mbox{$\YszM$}} scaling relation parameters affect the cluster redshift and tSZ \snr{} distributions, they can be constrained even in the absence of lensing masses albeit rather weakly \citep[accord][]{louis17}. 
For example, errors on cosmological parameters $\sigma(\summnu) = 60\ \mev$ and $\sigma(\wde) = 0.03$ both degrade by more than $\times2$ for pink without lensing compared to green curves with lensing mass calibration. 
Errors on virialization model parameters also degrade similarly by $\times 3$ or more without lensing mass information.

\subsubsection{Impact of high redshift clusters and redshift binning}
\label{sec_redshift_binning}
The constraints presented above have been derived by binning clusters at $z \ge 1.5$ in one massive high redshift bin. 
This is a conservative approach to take into account the difficulty of obtaining redshifts for distant clusters. 
We modify this choice using: case (i) an optimal setting with $\Delta z  = 0.1$ for all clusters; case (ii) a less conservative setting with $\Delta z = 0.1$ for $0.1 \le z < 2$ and $\Delta z = 1$ for $2 \le z \le 3$; and case (iii) a pessimistic setting by ignoring clusters at $z > 1.5$.

Case (i): We note $\times3.8$ better constraint on $\sigma(\vireffeta) = 0.0076$ compared to the baseline case $\sigma(\vireffeta) = 0.0228$ (see green curve in  Fig.~\ref{fig_constraints_s4_cmbhd}). 
Constraints on other virialization model parameters $\hsebias, \virslope, \virintercept$ also improve by 20-30\%. 
Similar improvements are seen for $h, \wde$ but this optimistic redshift binning scheme has negligible ($<10\%$) impact on 
$A_{s}$, $\summnu$, and $\omchsq$.

Case (ii): In this case, we see $\times3$ improvement on $\sigma(\vireffeta) = 0.0092$ and 15\% improvement on $\hsebias, \virslope, \virintercept$ compared to the baseline case but this setting has negligible ($<10\%$) impact on other parameters. 

Case (iii): Since $\vireffeta$ only affects clusters with $z \ge 2$, we do not constrain $\vireffeta$ with this pessimistic setting even though this is one of the main goals of this work.  
Nevertheless we perform this test to address the challenges in obtaining redshifts and understanding the survey selection for clusters at $z > 1.5$. 
We note up to $15\%$ degradation in constraints on cosmological parameters indicating that clusters with $z \le 1.5$ dominate cosmological constraints. 
The other virialization model parameter constraints, $\hsebias$ and $\virintercept$ worsen by 15\% while $\virslope$, controlling the redshift evolution of model 2, degrades by more than 60\%.

\subsubsection{Effect of $\sigma(\taure)$ prior}
\label{sec_tau_prior}
Here we check the effect of the \planck-like prior adopted in the forecasts above. 
Since a higher optical depth would suppress small-scale CMB anisotropies, $\taure$ has significant correlation with parameters like $h, \omchsq, \summnu, \wde$ and the choice of $\sigma(\taure)$ prior can affect other parameter constraints. 
Subsequently we check the effect of replacing \planck\ $\sigma(\taure) = 0.007$ with (i) no $\sigma(\taure)$ prior and (ii) a cosmic variance limited measurement $\sigma(\taure) = 0.002$, for example as expected from LiteBIRD or CORE satellites \citep{hazra18, divalentino18}. 
Even though this has an effect on CMB-only constraints on all the parameters listed above, we note significant effects only on $\summnu$ and $\omchsq$ with the joint CMB and cluster counts information. 

Removing $\taure$ prior degrades $\sigma(\summnu)$ by $\times 1.5$. 
With $\sigma(\taure) = 0.002$, $\sigma(\summnu)$ improves by $\times1.3-1.5$ from the three surveys. 
We obtain a roughly level of changes on $\sigma(\omchsq)$ with the two settings. 
Prior on $\taure$ does not affect virialization model parameter constraints. 

\subsection{Dependence on total survey time}
\label{sec_constraints_s4years}

\begin{figure}
\centering
\ifdefined\ApJsubmit
\includegraphics[width=0.48\textwidth, keepaspectratio]{clus_counts_S4wideyears.pdf}
\else
\includegraphics[width=0.48\textwidth, keepaspectratio]{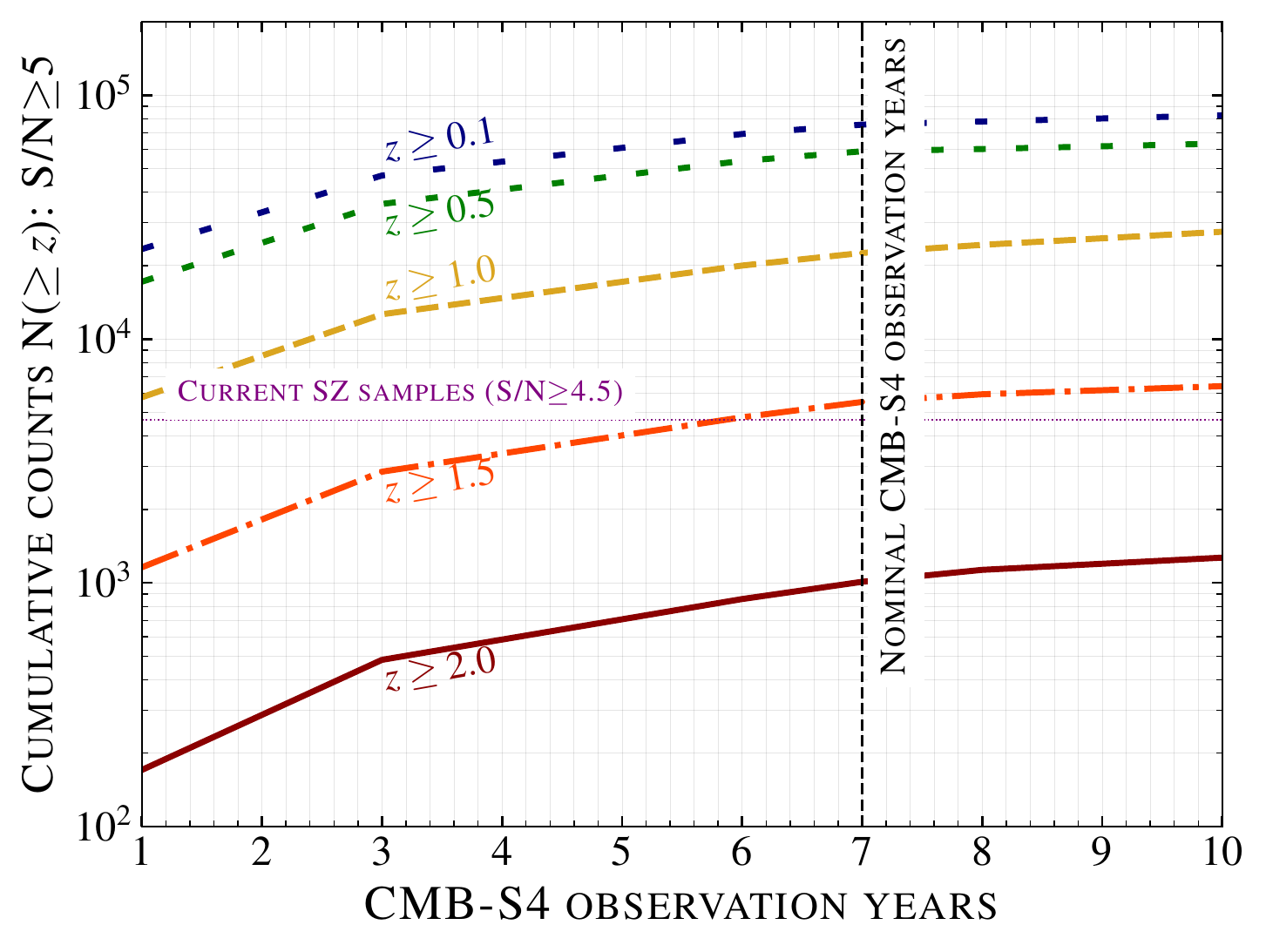}
\fi
\caption{
Cumulative cluster counts for \sfour{} shown as a function of number of observation years. The number of clusters in one year \sfour{} sample will surpass the currently available SZ cluster samples (purple dash-dotted line). 
The counts cannot be simply scaled from baseline observing period of 7 years because of residual foreground signals that dominate small-scales and hence have a major impact on high-$z$ clusters. 
For example, the actual $z \ge$ 2 clusters at the end of first year is $\times 2$ lower than what will be obtained using a simple noise scaling. 
}
\label{fig_s4wide_yearly_counts}
\end{figure}

Thus far, the forecasted cluster number counts and the cosmological constraints are obtained using noise levels in Table~\ref{tab_exp_specs} expected to be achieved at the end of survey periods. 
Given that \sfour{} is expected to start operations close to the end of this decade, these constraints may not be achieved until the middle of next decade. In this section, we check the dependence of cluster counts and cosmological constraints as a function of observing time for \sfour. 
For this test we simply scale the noise levels in each band by $\sqrt{N_{\rm years}/N_{\rm baseline}}$ where \mbox{$N_{\rm baseline}$ = 7 years}. 
We go from 1 to 10 years. 
Note that this simple scaling assumes full deployment at the start of the survey which could be unrealistic and the noise scaling may be slightly more complicated for the first few years in reality. 

\begin{figure}
\centering
\ifdefined\ApJsubmit
\includegraphics[width=0.48\textwidth, keepaspectratio]{cosmoconstraints_S4wideyears.pdf}
\else
\includegraphics[width=0.48\textwidth, keepaspectratio]{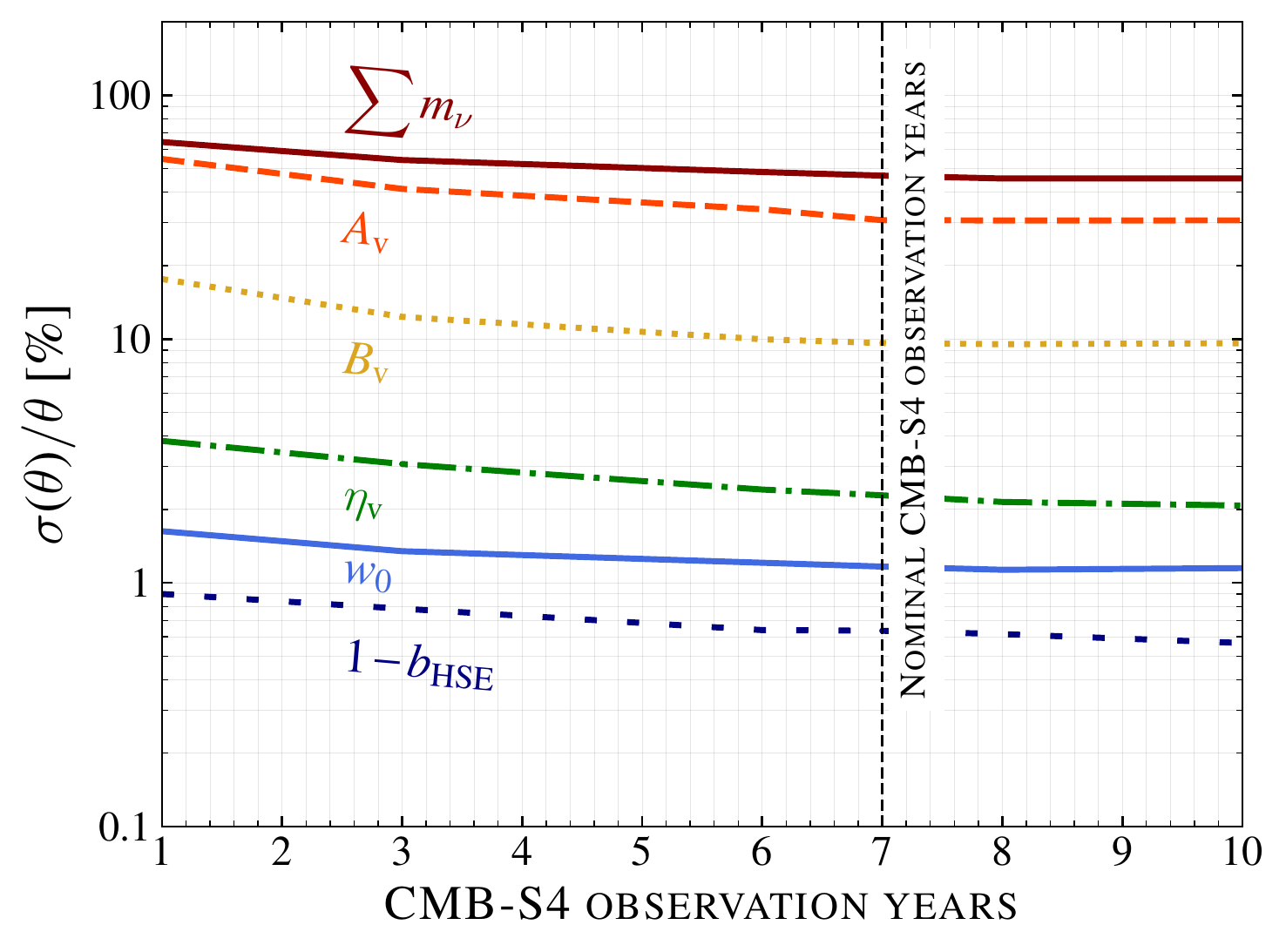}
\fi
\caption{Relative cosmological constraints that can be obtained by combining clusters with primary CMB as a function of \sfour{} observing years. 
Our results indicate that \sfour{} can return remarkable cosmological constraints compared to current limits even in its first few years of observation. 
\sfour, at the end of first few years of observations, can also place compelling constraints on the virialization mechanism of high redshift clusters that is currently unconstrained.
}
\label{fig_s4wide_yearly_constraints}
\end{figure}

In Fig.~\ref{fig_s4wide_yearly_counts}, we show the cumulative cluster counts for clusters above multiple redshifts as a function of number of \sfour{} observation years. 
It is evident from the figure that \sfour{} cluster sample, with $\sim 20,000$ clusters, will surpass the current SZ samples from ACT, \planck, and SPT (purple dotted line) even at the end of the first year of observation. 
In fact, at the end of first year \sfour{} sample will have close to 5000 clusters at $z \ge 1$, similar to the total number of current SZ clusters at all redshifts. 
It is also worth noting that we use a \snr{} threshold of 5 for \sfour{} while current SZ samples shown here use 4.5. 

The number of clusters expected at the end of each year cannot be obtained from baseline results (7 years) using a simple noise scaling because of residual foregrounds in the Compton-$y$ maps. 
Since the residual foregrounds (mostly CIB) dominate small scales, the impact of residual foregrounds is more important for high redshift clusters that span a smaller angular extent on the sky. 
For example, scaling \sfour{} clusters at $z \ge 2$ from Table~\ref{tab_cluster_counts} should return $N_{\rm clus}^{\rm 1 year}(z \ge 2) = \sqrt{7}(992) = 375$ clusters while we have $\lesssim 200$ clusters ($\times 2$ lower) at the end of year 1 in Fig.~\ref{fig_s4wide_yearly_counts}. 

Cosmological constraints from clusters and primary CMB as a function of \sfour{} observation years are presented in Fig.~\ref{fig_s4wide_yearly_constraints}. 
Similar to cluster forecasts, noise levels in each band are scaled from Table~\ref{tab_exp_specs} for each year to obtain the CMB Fisher matrix. 
The improvement in constraints is not dramatic as a function of observing years and this is primarily because of different degeneracy directions in the parameter space probed by clusters and primary CMB. 
However, note that cluster cosmology is not the only science driver for CMB-S4. 
It has a broad range of science goals including the measurement of light relic density and the production legacy catalogues in mm/sub-mm wavelengths that requires the proposed \mbox{$N_{\rm baseline}$ = 7 years} \citep{cmbs4collab19} to produce wide and deep CMB maps. 
With $\sim 200 (400)$ clusters at $z \ge 2$ at the end of year 1 (3), we find the \sfour{} can make a giant leap towards understanding the gastrophysics and the onset of virialization mechanism of high redshift clusters, which are completely unconstrained currently. 

\subsection{Effect of cluster correlated foreground signals}
\label{sec_clus_corr_fg}
In our baseline approach we ignored cluster correlated foreground signals namely the cluster kSZ signal and emission from DSFG and RG within clusters. 
For clusters close to detection limits for all the three surveys considered here, cluster kSZ signals are less important as they are expected to be much smaller than the tSZ signal. 
Moreover, kSZ can be both positive or negative depending on the direction of the radial motion and hence only acts as an additional source of variance in our analysis. 
On the other hand, emission from DSFG and radio galaxies, since they are always positive, can fill in the tSZ decrements thereby potentially contaminating tSZ measurements. 
While the presence of DSFG signals within clusters have been identified in \planck{} clusters \citep{planck16clusterdustsed}, \citet{melin18} reported that signals from DSFGs degrade the completeness of \planck{} cluster catalog by $\lesssim 10\%$ and showed that this contamination has negligible impact on cosmological parameter inference.
However, \planck{} cluster sample deals with massive low redshift clusters where the star formation has been observed to be highly suppressed \citep{popesso15}. 
Furthermore, the cluster tSZ signal goes as $M^{5/3}$ while DSFG signals are roughly linear. 
As a result, DSFG contamination may be insignificant for \planck{} clusters. 
In this work, we are particularly interested in low mass ($\mvir \lesssim 10^{14} \msol$) and high redshift $ z \ge 1.5$ clusters, near the peak of the cosmic star formation history, to constrain cluster gastrophysics and hence dust contamination might be potentially important. 

We check the impact of cluster correlated signals using Websky \citep{stein20} and MDPL2 \citep{omori21prep} simulations. 
Websky\footnote{\url{https://mocks.cita.utoronto.ca/websky}} simulations are publicly available while MDPL2\footnote{\url{http://behroozi.users.hpc.arizona.edu/MDPL2/hlists/}} is currently under preparation and obtained using private communication.
For this test, it is important that the correlation between tSZ and CIB signals in Websky and MDPL2 simulations is in agreement with the measurements reported in the literature. 
Defining the correlation coefficient between the two as $\rho_{\rm tSZ \times CIB} = \dfrac{C_{\ell}^{\rm tSZ \times CIB}}{\sqrt{C_{\ell}^{\rm tSZ \times tSZ} C_{\ell}^{\rm CIB \times CIB}}}$, the value for Websky (MDPL2) at $\ell=3000$ is $\sim 0.25\ (0.17)$. 
These are in reasonable agreement with the values reported by SPT $0.2 \pm 0.12$ \citep{george15} and \planck{} $0.18 \pm 0.07$ \citep{planck16tszcib}.
To this end, we extract \boxsize{} cutouts of kSZ and DSFG emissions around haloes in the mass and redshift grid used for \snr{} calculations: \mbox{${\rm log}\mvir \in [13,15.4]\ \msol$} with \mbox{${\rm log}\Delta M = 0.1\ \msol$} and $z \in [0.1, 3]$ with $\Delta z = 0.1$. 
Before extraction, we mask sources with flux at 150 GHz above $S_{150} \sim 6$ mJy. 
We also apply a frequency dependent scaling factor for Websky DSFG signals to match SPT measurements. 
At 150 GHz, this map scaling factor is 0.75 to match Websky to SPT $D_{\ell_{3000}}^{150} = 12 \mu K^{2}$ \citep{reichardt21}. 
No such scaling was applied to MDPL2 simulations.
In both cases, we pick 100 cutouts for every point in the $\mvir, z$ grid. 
Due to the availability of a single Websky/MDPL2 mock sky realization, the number of kSZ/CIB signals available for this test reduces significantly for clusters with mass $\mvir \gtrsim 3 \munits$ at $z \gtrsim 2$. 
Hence, we limit this test to clusters below this mass and redshift range. 
\%refresponse{
We inject the cluster-correlated signals from Websky/MDPL2 into our simulations in all the frequency bands along with cluster tSZ signal, experimental noise, CMB and other astrophysical foregrounds described in \S\ref{sec_sims_overview} which are then passed through the ILC pipeline.

For high redshift ($z \ge 1$) clusters near the detection limit, we note that DSFG within clusters shift the tSZ-based cluster masses slightly lower. 
However, the bias is sub-dominant compared to statistical uncertainties at roughly $\le 0.2 \sigma$ level for all the three surveys. 
The bias is almost zero for low redshift clusters. 
As expected, cluster kSZ signals have negligible impact on the recovered tSZ signals. 

Both Websky and MDPL2 do not contain emission from RG within clusters. 
Subsequently, we choose an extremely conservative test to assess the impact of RG within clusters on the recovered tSZ signals. 
We inject a constant flux of $S_{150}=$ 0.1 mJy or 0.5 mJy for all clusters where the latter roughly matches the point source sensitivity ($3\sigma$) at 90 GHz for CMB-S4 survey \citep{cmbs4collab19}. 
The signal is scaled to other bands assuming a spectral index $\alpha_{\rm radio} = -0.6$ \citep{everett20}.
This test is limited to \sfour{} only.
We find that a \mbox{$S_{150}=$ 0.1 mJy (0.5 mJy)} can bias tSZ measurements low by $0.1-0.2\sigma\ (\gtrsim 1\sigma)$. 
While a $1\sigma$ systematic error is large, we note that our model for RG is unrealistic and hence our results should only be interpreted as an upper limit of the systematic error. 

Our simple RG model can be potentially replaced using the Websky simulations which are currently being upgraded to include RG signals correlated with the underlying dark matter. 
Similarly the limitation due to the smaller number of Websky or MDPL2 haloes at high mass end can be addressed using multiple realizations of the millimetre wave sky, as released recently by \citet{han21} using deep learning techniques for example.
We leave these detailed studies for a future work. 

\subsubsection{Impact on sample purity due to point sources}
\label{sec_sample_purity}
Given that point sources in the maps can be misclassified as clusters, we check the effect of point sources using 100 noise-only simulations. 
The simulations for this test include astrophysical foregrounds, CMB, experimental noise and point source signals. 
Cluster tSZ signal is ignored here. We model the point source signals in three different ways. 

In the first case, we add the cluster-correlated DSFG signals in each mass and redshift bin using Websky simulations as explained above. 
We obtain zero false detections which is consistent with a negligible systematic bias from dusty sources estimated above in \S\ref{sec_clus_corr_fg}. 
In a similar spirit, we also check the effect of random point sources in the maps. In this case, we inject point source signals with fluxes $S_{150} \in [0.5, 1, 2, 3]$ mJy which are then scaled to other bands using a power-law relation with a spectral index $\alpha_{\rm dust} = 3.2$ and $\alpha_{\rm radio} = -0.6$ to represent DSFG and radio point source signals \citep{george15}. 
Again, we do not see any false detections from DSGFs. 
This indicates that DSFGs do not show up as $\ge 5\sigma$ positive peaks in the Compton-$y$ maps.
For radio point sources, we find that the sources with flux $S_{150} \ge 2$ mJy can be potentially problematic. 
However, these radio point sources show up as negative peaks in the Compton-$y$ maps and hence will not be classified as clusters.


\section{Conclusion}
\label{sec_conclusion}

We forecasted the number of galaxy clusters that can detected using future CMB surveys namely \sfour, \sfourdeep, and \cmbhd.
Our forecasts used realistic simulations that include signals from galactic and extragalactic signals along with atmospheric and instrumental noise components. 
In the baseline footprint with $\fsky = \fskywideclean$, \sfour{} can detect close to 75,000 clusters and \cmbhd{} sample will contain $\times 3$ more clusters. 
The smaller but deeper \sfourdeep{} survey can detect $\sim 14,000$ clusters. 
Of these, 6000 (1500) will be at $z \ge 1.5$ and 1000 (350) clusters will be at $z \ge 2$ in the \sfour{} (\sfourdeep) cluster sample. 
The number of $z \ge 2$ clusters is an order of magnitude higher for the \cmbhd{} experiment. 
Including regions close to the galactic plane ($\fsky = \fskywidedirty$) increases the sample size by roughly 20\%. 

The residual foreground signals, CIB in particular, dominates the small-scale variance in the Compton-$y$ maps for CMB-S4. 
For \cmbhd, the variance from CIB is $\times17$ lower at 150 GHz due to efficient subtraction of dusty galaxy sources \citep{sehgal19}. 
The CIB subtraction along with a $\times 5$ smaller beam are the reasons for a much larger cluster sample from \cmbhd. 
Given the importance of residual CIB signals and the tSZ $\times$ CIB correlation, we checked the systematic errors in the recovered cluster tSZ signals due to emission from galaxies within clusters. 
The effect of dusty star forming galaxies was studied using Websky/MDPL2 simulations while for radio galaxies we use a simple constant flux model for all clusters. 
Our results indicate that systematic error due to the presence of dusty galaxies is much smaller than the statistical error $\le 0.2\sigma$ but having a constant radio galaxy signal with $S_{150} = 0.5$ mJy can introduce $\sim 1\sigma$ bias. 
The models used for galactic foregrounds and RG are basic and must be extended further. Nevertheless, the tests we performed to assess their contamination on the recovered tSZ signals is important for future SZ surveys.

We used CMB-cluster lensing signal from both temperature and polarization to calibrate the cluster tSZ-mass scaling relation. 
We have ignored weak lensing informations from optical surveys in this work and note that including them can further improve the constraining power as well as act as an important systematic check for CMB-cluster lensing based mass estimates.
The internally calibrated cluster counts was combined with primary CMB (TT/EE/TE) spectra to derive cosmological constraints assuming a two parameter extension to the standard model of cosmology ($\lcdm + \summnu + \wde$). 
We show that the constraints on dark energy equation of state $\sigma(\wde)$ parameter can be between $1-2\%$ for \sfour/\sfourdeep{} and sub-percent for \cmbhd. 
Similarly, the sum of neutrino masses $\summnu$ can detected at $\gtrsim 2.5-4.5\sigma$ by both CMB-S4 and \cmbhd{} surveys assuming a normal hierarchy lower limit of $60\ \mev$. 
We also assess the importance of combining cluster counts with primary CMB, significance of CMB-cluster lensing, choice of $\sigma(\taure)$ prior, effect of different redshift binning and dependence of our result on the total survey time. 

Besides cosmology and scaling relation constraints, we also study the evolution of the ICM using two models. 
In the first case, we model cluster virialization in Eq.(\ref{eq_vir_model_1}) using the standard HSE bias parameter and a virialization efficiency parameter $\vireffeta$ that linear scales the tSZ signal from high redshift \mbox{$z \ge 2$} clusters. 
We find $\sigma(\vireffeta) = 0.00471, 0.0228, 0.0339$ from \cmbhd, \sfour, and \sfourdeep{} indicating that the mean deviation in thermal energy of $z \ge 2$ clusters from their low redshift counterparts can be constrained to roughly $2-4\%$ by CMB-S4 and $<1\%$ by \cmbhd{} experiments. All the three surveys can place sub-percent constraints on the HSE bias parameter. 
Our second model in Eq.(\ref{eq_vir_model_2}) is more physically motivated and calibrated using Omega500 hydrodynamical cosmological simulations. 
In this case, CMB-S4 can provide $\sim 4\%$ constraints on $\virintercept$ and $\sim 33\%$ on $\virslope$ that controls the redshift evolution of the virialization mechanism. 
\cmbhd{} improves the constraining power by more than $\times3$.  
This work represents a key step towards understanding the selection function of high redshift clusters and the evolution of ICM using current and future CMB SZ surveys like AdvACT \citep{henderson16}, \cmbhd{} \citep{sehgal19}, CMB-S4 \citep{cmbs4collab19}, SPT-3G \citep{benson14, bender18} and SO \citep{SO18}. 
The binned cluster counts $\binnedcluscounts$, Fisher matrices, and other associated products can be downloaded from this \href{https://github.com/sriniraghunathan/tSZ_cluster_forecasts}{link$^{\text{\faGithub}}$}.


\section*{Acknowledgments}
We thank the entire CMB-S4 collaboration\footnote{\url{https://people.cmb-s4.org/public/showdir.php}} for helpful comments and suggestions throughout the course of this work. 
We further thank Neelima Sehgal for feedback on the manuscript; Sebastian Bocquet and Nikhel Gupta for useful discussions; and Yuuki Omori for providing access to MDPL2 simulations. 
Finally, we thank the anonymous referee for useful suggestions that helped in shaping this manuscript better.
SR is supported by the Illinois Survey Science Fellowship from the Center for AstroPhysical Surveys at the National Center for Supercomputing Applications.
SR and NW acknowledge support from NSF grants AST-1716965 and CSSI-1835865.
SR, NW, GH, and JV acknowledge support from NSF grant OPP-1852617.
DN and HA acknowledge support from the facilities and staff of the Yale Center for Research Computing.
NB acknowledges support from NSF grant AST-1910021.
EP is supported by NASA grant 80NSSC18K0403 and the Simons Foundation award number 615662; as well as  NSF grant AST-1910678. 
CR acknowledges support from the Australian Research Council's Discovery Projects scheme (DP200101068).
JV acknowledges support from NSF under grants AST-1715213 and AST-1716127.
This work used computational and storage services associated with the Hoffman2 Shared Cluster provided by UCLA Institute for Digital Research and Education's Research Technology Group.

\appendix
\restartappendixnumbering
\section{A physically-motivated cluster virialization model}
\label{sec_cluster_vir_model2}

Galaxy clusters are dynamically active objects and generally out of HSE due to mergers and mass accretion processes. The lack of virialization is characterized by a non-thermal pressure fraction, $P_{\rm th}/P_{\rm tot}$, which quantifies the fraction of energy densities in unvirialized bulk and turbulent gas motions compared to the total pressure, $P_{\rm tot}=P_{\rm th}+P\nth$ \citep[e.g.,][]{lau09,Battaglia12,nelson14a,Shi14,Yu15}. In presence of the non-thermal pressure, the total pressure is given by the sum of thermal and non-thermal pressure: $P_{\rm tot}=P_{\rm th}+P\nth$, which in turn provides the pressure support against the gravitational collapse.  Since the tSZ effect is sensitive to only the thermal pressure component $Y_{\rm tSZ} = Y_{\rm tot}-Y\nth$, the observed integrated tSZ signal is reduced by
\begin{equation}
    \frac{Y\nth}{Y_{\rm tot}} = 1-[\vireffeta(z)(1-\hsebias)^{\alphay}],
\end{equation}
which represents the combination of the lack of virialization and HSE mass bias from Eq.(\ref{eq_Ysz_mass}) and $Y$ is the integrated pressure of each component within the sphere of $\rvir$ following Eq.(\ref{eq_compton_y}). 
Note that the non-thermal pressure is one of the dominant sources of systematic uncertainties in the HSE mass bias \citep[e.g.,][]{Nagai07b,Lau13,Shi16,Biffi16,Angelinelli20}. 

\begin{wrapfigure}{rt}{0.45\textwidth}
\vskip -15pt
\ifdefined\ApJsubmit
\includegraphics[trim=0.1in 0.in 0in 0.0in,clip=true, width=0.45\textwidth]{Ynth_z_evolution.pdf}
\else
\includegraphics[trim=0.1in 0.in 0in 0.0in,clip=true, width=0.45\textwidth]{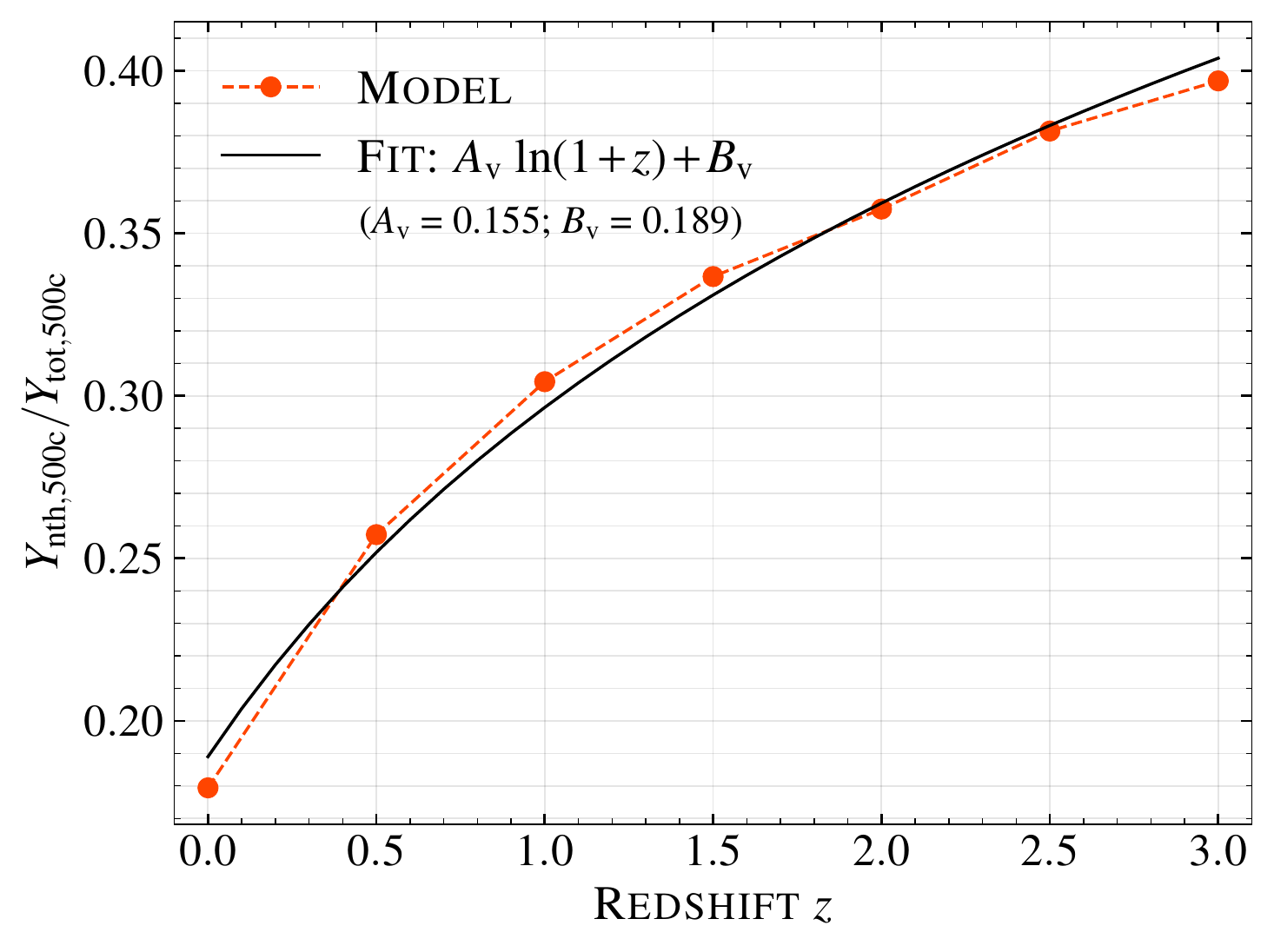}
\fi
\caption{The ratio of the non-thermal pressure fraction, $Y{\nth}/Y_{\rm tot}$, enclosed within the projected aperture radius of $\rvir$ for a $\mvir=10^{14}\ \msol/h$ galaxy cluster as a function of redshift $z$. The model (orange) predicts that the non-thermal pressure increases as a function of redshift due to the enhanced mass accretion rate in the early Universe. We propose a fitting function (black solid), Eq.(\ref{eq:yratio_fit}), which describes the redshift evolution out to $z\approx 3$.}
\vskip -10pt
\label{fig:Yratio}
\end{wrapfigure}

We compute the impact of the non-thermal pressure on the \mbox{$\YszM$} relation of high redshift clusters using the model presented in \citet{Green20}. First, we use the analytical model of \citet{Shi14} to compute the evolution of non-thermal pressure, by solving 
\begin{equation}\label{eqn:evolve}
    \frac{\rm d\sigma\nth^2}{\rm dt} = -\frac{\sigma\nth^2}{t_\mathrm{dis}} + \eta \frac{\rm d \sigma_\mathrm{tot}^2}{\rm dt},
\end{equation}
where $\sigma\nth^2 = P\nth/\rho$ denotes the velocity dispersion due to non-thermal random motion, $t_{\rm dis}$ is the dissipation time of the turbulence scale $t_{\rm dis}$, $\sigma_{\rm tot}^2 = P_{\rm tot}/\rho$ is the total velocity dispersion and $\eta$ is the fraction of energy accreted that are injected into turbulence motion. 
Due to the cosmic mass accretion process, the total velocity dispersion increases over time. The turbulence decays into the thermal energy over the dissipation time scale $t_{\rm dis}$, which is proportional to the eddy turn-over time of the largest eddies, which is in turn proportional to the local orbital time, $t_\mathrm{dis}(r) = \beta t_\mathrm{orb}(r) / 2$. The model has been calibrated using Omega500 hydrodynamical cosmological simulations \citep{nelson14b}, yielding the best fit parameters of $\beta = 1$ and $\eta=0.7$ \citep{Shi15}. Given a cluster with mass $\mvir$, we generate the average mass accretion history $M(t)$ and concentration $c(t)$ of the cluster following \citet{vdBosch14} and \citet{Zhao09}, respectively. For the total pressure profiles, we use the KS01 \citep{komatsu01} model, which is based on a polytropic gas in HSE with the NFW profile. 

Figure~\ref{fig:Yratio} shows the fraction of $Y(<R_{\rm 500c})$ signal in non-thermal pressure, $Y{\nth}/Y_{\rm tot}$ as a function of redshift. 
For a constant mass $\mvir=10^{14} \msol/h$, the model predicts that the fraction of $Y$ signal in non-thermal pressure can evolve from $20\%$ at $z=0$ to $40\%$ at $z=3$, indicating strong redshift dependence. 
The model predicts the enhancement in the non-thermal pressure fraction toward high-redshift due to the enhanced mass accretion rate in the early universe \citep{Green20}.
Our result suggests that the evolution of $Y{\nth}/Y_{\rm tot}$ at $\mvir =10^{14}\ \msol/h$ can be described by a simple function: 
\begin{equation}\label{eq:yratio_fit}
    \frac{Y{\nth}}{Y_{\rm tot}} = \virslope \ln(1+z) + \virintercept,
\end{equation}
where $\virslope$ and $\virintercept$ are calibrated to $\virslopeval$ and $\virinterceptval$. 

\bibliographystyle{aasjournal}
\bibliography{ICM}

\end{document}